\documentclass[11pt,a4paper]{article}
\pdfoutput=1
\usepackage{jcappub}
\usepackage{aas_macros_JCAP}
\usepackage{caption}
\usepackage{subcaption}
\usepackage{blindtext}

\newcommand{\hMpc}{h^{-1}{\rm Mpc}}

\newcommand{\hGpc}{h^{-1}{\rm Gpc}}
\newcommand{\hMsun}{h^{-1}M_{\odot}}
\newcommand{\void}{\mathrm{v}}
\newcommand{\halo}{\mathrm{h}}

\newcommand{\tracer}{\mathrm{t}}
\newcommand{\matter}{\mathrm{m}}

\newcommand{\sumnu}{\sum m_{\nu}}
\newcommand{\coreDens}{\hat{n}_\mathrm{c}}

\newcommand{\AP}[1]{#1}
\newcommand{\EM}[1]{#1}
\newcommand{\CK}[1]{#1}

\newcommand{\CC}[1]{#1}
\newcommand{\REF}[1]{#1}

\title{The bias of cosmic voids in the presence of massive neutrinos}

\author[a]{Nico Schuster,}
\author[a,b]{Nico Hamaus,}
\author[c]{Alice Pisani,}
\author[d,e,f]{Carmelita Carbone,}
\author[c]{Christina D. Kreisch,}
\author[a,b]{Giorgia Pollina,}
\author[a,b,g]{and \\Jochen Weller}

\affiliation[a]{Universit\"ats-Sternwarte M\"unchen, Fakult\"at f\"ur Physik, Ludwig-Maximilians Universit\"at, Scheinerstr. 1, D-81679 M\"unchen, Germany}
\affiliation[b]{Excellence Cluster Origins, Bolzmannstr. 2, D-85748 Garching, Germany}
\affiliation[c]{Princeton University, Department of Astrophysical Sciences, 4 Ivy Lane, Princeton, NJ 08544, USA}
\affiliation[d]{INAF -- Istituto di Astrofisica Spaziale e Fisica Cosmica Milano, via Alfonso Corti 12, I-20133 Milano, Italy}
\affiliation[e]{INFN, Sezione di Milano, via Giovanni Celoria 16, I-20133 Milano, Italy}
\affiliation[f]{Universit\`a degli studi di Milano, Dipartimento di Fisica, via Giovanni Celoria 16, I-20133 Milano, Italy}
\affiliation[g]{Max Planck Institute for Extraterrestrial Physics, Giessenbachstr. 1, D-85748 Garching, Germany}

\emailAdd{nschuster@usm.lmu.de}
\emailAdd{n.hamaus@physik.lmu.de}

\abstract{
Cosmic voids offer an extraordinary opportunity to study the effects of massive neutrinos on cosmological scales. Because they are freely streaming, neutrinos can penetrate the interior of voids more easily than cold dark matter or baryons, which makes their relative contribution to the mass budget in voids much higher than elsewhere in the Universe. In simulations it has recently been shown how various characteristics of voids in the matter distribution are affected by neutrinos, such as their abundance, density profiles, dynamics, and clustering properties. However, the tracers used to identify voids in observations (e.g., galaxies or halos) are affected by neutrinos as well, and isolating the unique neutrino signatures inherent to voids becomes more difficult. In this paper we make use of the DEMNUni suite of simulations to investigate the clustering bias of voids in Fourier space \AP{as a function of their core density and compensation}. We find a clear dependence on the sum of neutrino masses that remains significant even for void statistics extracted from halos. In particular, we observe that the amplitude of the linear void bias increases with neutrino mass for voids defined in dark matter, whereas this trend gets reversed and slightly attenuated when measuring the relative void-halo bias using voids identified in the halo distribution. Finally, we argue how the original behaviour can be restored when considering observations of the total matter distribution (e.g. via weak lensing), and comment on scale-dependent effects in the void bias that may provide additional information on neutrinos in the future.
}

\date{\today}

\keywords{cosmological parameters from LSS, neutrino masses from cosmology, cosmological simulations, cosmic web, galaxy clustering, power spectrum, redshift surveys}

\arxivnumber{1905.00436}

\begin{document}
\maketitle

\section{Introduction\label{sec:intro}}
Oscillation experiments have shown that neutrinos must possess mass and that at least one of the three species weighs $m_\nu>0.06$eV \cite{Fukuda1998}. Despite their extremely low masses, neutrinos play an important role in the form of \emph{hot} dark matter for the Universe as a whole. While neutrinos were relativistic at very early times, they have cooled down and became non-relativistic in the course of cosmic evolution, which means they can cluster gravitationally and interact with the remaining mass distribution in a nonlinear fashion. Their signatures may be observed both in the cosmic microwave background (CMB) and the large-scale structure (LSS) \cite[e.g.,][]{Lesgourgues2006,Villaescusa2018,Liu2018}. The heavier they are, the more they interact gravitationally and the smaller is their \emph{free-streaming} length. The latter determines the transition scale below which the spatial fluctuations in their distribution become suppressed due to thermal pressure. In the total matter power spectrum this causes a characteristic suppression that affects the abundance of galaxy clusters and the two-point statistics of galaxies~\cite{Roncarelli2015}. There are first indications for the imprints of neutrinos in galaxy survey data when combined with the CMB \cite{Beutler2014,Baumann2018}, but the results are still controversial and need further confirmation \cite[e.g.,][]{Costanzi2014,Alam2017}. For example, the latest observations of the Lyman-$\alpha$ forest, as well as the latest CMB data from Planck, constrain the sum of all neutrino masses to $\sum m_\nu<0.12$eV \cite{Palanque-Delabrouille2015,Planck2018Parameters}, leaving only a small open window to the lower limit of $0.06$eV. Hence, the prospects for a successful detection of neutrino masses are in immediate reach with current and future data sets~\cite[e.g.,][]{BOSS,DES,EUCLID,LSST,WFIRST}.

\REF{Instead of using the denser part of the cosmic web for the search of massive neutrino signatures we focus on cosmic voids, the large and under-dense regions of the large-scale structure. These voids are the largest structures in the cosmos with sizes generally ranging from around tens to a hundred $\hMpc$, filling most of the volume of the Universe. Within these regions matter is sparsely distributed, with central densities in between $10\%$ and $50\%$ of the mean cosmic density. The dynamics of these regions are simpler compared to overdense structures and they are dominated by the outward bulk flows while having undergone minimal virialization~\cite{Weygaert1993,Gottloeber2003,Shandarin2011,AragonCalvo2013,Hamaus2014c,Falck2015,Ramachandra2017}.
The hierarchical substructure of voids even resembles a miniature cosmic web, however with a different and clearly lower mean density~\cite{Goldberg2004}. The existence of these voids has been a long-time prediction of the cosmological standard model~\cite{Hausman1983}, and their first observation dates back almost 40 years to galaxy redshift surveys of the 1970s~\cite{Gregory1978,Kirshner1981}. Only recently systematic void studies have become feasible. This is due to large cosmological simulations with the capability to predict the matter distribution with high accuracy, as well as due to deep galaxy surveys covering significant fractions of the sky.}

Cosmic voids provide ideal environments to search for signatures from massive neutrinos, because the latter can escape the surrounding gravitational potential wells more efficiently than other particles, such as cold dark matter (CDM) or baryons. Thus, the relative neutrino abundance in voids can be much higher than elsewhere in the Universe. In addition, the typical extent of voids coincides with the expected neutrino free-streaming scale \CC{at late times}~\CK{(see~\cite{Kreisch2018} for the calculation)}, which makes voids a prime target to look for neutrino effects. Similar arguments apply to other potentially interesting particle species, such as \emph{warm} dark matter or axions, \CK{which also exhibit a characteristic free-streaming scale. For example, reference~\cite{Yang2015} analyzed warm dark matter cosmologies with numerical simulations, finding shallower density profiles for voids}. At the same time, observational results from cosmic voids have rapidly accumulated in the recent years, which lead to a plethora of new cosmological signals and discoveries, such as imprints from weak lensing (WL)~\cite[e.g.,][]{Melchior2014,Clampitt2015,SanchezC2017,Fang2019}, the integrated Sachs-Wolfe (ISW) effect~\cite[e.g.,][]{Granett2008,Nadathur2016,Kovacs2017,Kovacs2019}, the Sunyaev-Zel'dovich (SZ) effect~\cite{Alonso2018}, baryon acoustic oscillations (BAO)~\cite{Kitaura2016}, the Alcock-Paczy\'nski (AP) effect~\cite[e.g.,][]{Sutter2012b,Hamaus2016,Mao2017} and redshift-space distortions (RSD)~\cite[e.g.,][]{Hamaus2015,Achitouv2017,Hawken2017,Hamaus2017} that have been found in void data.

\REF{Only recently dedicated simulation studies have uncovered some of the effects that neutrinos have on cosmic voids, although the latter have already been of interest as cosmological probes to investigate differences between CDM- vs. neutrino-dominated cosmologies before~\cite[e.g.,][]{Zeldovich1982,Davis1985}. The aforementioned recent studies, which treat neutrinos as a sub-dominant component in the cosmic energy budget, generally find} shallower density profiles and attenuated velocity profiles for voids in massive neutrino cosmologies~\cite{Massara2015}, \CK{in agreement} with the naive expectations mentioned above. The existence of neutrinos also tends to suppress the abundance of small voids, while boosting the number of large voids found in the halo distribution~\cite{Massara2015,Kreisch2018}. Moreover, \CK{neutrinos enhance the two-point clustering amplitude of voids in a non-trivial scale-dependent way}~\cite{Banerjee2016,Kreisch2018}. All of these smoking-gun signatures, \CK{however}, depend on the choice of tracer field used to identify the voids in~\CK{\cite{Pollina2017,Kreisch2018}}. Ideally this would be the total matter density field, which results in the strongest neutrino imprints on voids. Unfortunately, we cannot create a three-dimensional map of the dark matter distribution with the radial resolution needed to see voids. However, galaxy surveys provide us with a 3D distribution of biased tracers of the \CC{matter} density field, and we can use those to identify voids~\cite[e.g.,][]{vide_general_paper}. In this case, some effects that neutrinos have on voids can be mitigated~\cite{Massara2015,Kreisch2018}, because the distribution of biased tracers themselves is affected by neutrinos. Similar conclusions have been drawn in the context of modified gravity~\cite{Cai2015,Cautun2018} and coupled dark energy models~\cite{Pollina2016}.

The goal of this paper is to study in detail how void abundance and void clustering are affected by massive neutrinos, and how the tracer type in the identification of voids matters in this respect. \EM{}\CK{}\AP{In particular, we focus on the dependence of these statistics on key void properties that we expect to be most sensitive to neutrinos, such as the core density and the compensation of voids.} In section \ref{sec:theory} we briefly summarize some theoretical aspects of how to describe neutrinos in large-scale structure. Section \ref{sec:data} provides details on the simulation suite and the method of void finding we used. In section \ref{sec:analysis} we present our analysis results, with a model discussion in section \ref{sec:discussion} and final conclusions in section \ref{sec:conclusion}.

\section{Theory \label{sec:theory}}

\subsection{Massive neutrinos in cosmology \label{subsec:neutrino_cosmology}}

In the early Universe massive neutrinos behave as relativistic particles and are considered as hot dark matter. Only at later times they become non-relativistic and the transition happens when the temperature of neutrinos drops below their mass. The redshift for when this happens is \CK{$1 + z_{\mathrm{nr}} \simeq 1890\,( m_{\nu}/1 \mathrm{eV})$}~\cite{neutrino_redshift}. This relation holds for each neutrino species individually and depends only on the mass of each species. After this time the non-relativistic massive neutrinos contribute with $\Omega_{\nu}$ to the energy density budget as dark matter and the total matter energy density is given by \CK{$\Omega_{\mathrm{m}} = \Omega_{\mathrm{cdm}} + \Omega_\mathrm{b} + \Omega_{\nu}$}. Today the value of $\Omega_{\nu}$ is proportional to the sum of all neutrino masses $\sumnu$, in units of the critical density~\cite{neutrino_omega_mass}:
\begin{equation}
    \Omega_{\nu} = \frac{\sumnu}{93.14 \, h^2 \, \mathrm{eV}}.
\end{equation}
Since we are interested in \CC{neutrino} effects on scales much larger than the Jeans length of the baryons, we define $\Omega_\mathrm{c} \equiv \Omega_\mathrm{b} +  \Omega_{\mathrm{cdm}}$ as the density of all cold matter and treat it as a single component. At the perturbative \CC{linear} level massive neutrinos have an influence on fluctuations in the matter density and therefore also influence the total matter power spectrum. With the neutrino fraction $f_{\nu} \equiv \Omega_{\nu} / \Omega_\mathrm{m} $ the total matter power spectrum reads
\begin{equation}
    P_{\mathrm{mm}} = \left( 1 - f_{\nu} \right)^2 P_{\mathrm{cc}} + 2 f_{\nu} \left( 1 - f_{\nu} \right) P_{\mathrm{c}\nu} + f_{\nu}^2 \, P_{\nu \nu},
\end{equation}
where $P_{\mathrm{cc}}$ is the cold matter power spectrum, $P_{\nu \nu}$ is the power spectrum of neutrinos and $P_{\mathrm{c}\nu}$ is the cold matter and neutrino cross-power spectrum. \CC{For the nonlinear regime one can apply a mapping to $P_{\mathrm{mm}}$ using the nonlinear $P_{\mathrm{cc}}$ only~\cite{DEMNUni_clustering_paper}, but in this paper we restrict our analysis to the linear regime.}

The average distance that neutrinos travel over the age of the Universe depends on their thermal velocity and therefore their mass. This distance is referred to as the free-streaming length $\lambda_{\mathrm{FS}}$. After neutrinos become non-relativistic, their \emph{comoving} free-streaming length $\lambda_{\mathrm{FS}}/a$ reaches a maximum in the matter dominated epoch of the Universe and decreases afterwards. \CC{This maximum} is related to a minimal comoving wavenumber \cite{neutrino_omega_mass}
\begin{equation}
    k_{\mathrm{nr}} \simeq 0.018 \, \Omega_{\mathrm{m}}^{1/2 } \left( \frac{m_{\nu}}{1 \mathrm{eV}} \right)^{1/2} h\, \mathrm{Mpc}^{-1}.
\end{equation}
Scales smaller than $1/k_{\mathrm{nr}} $ are affected by the presence of massive neutrinos and we can write in first approximation \cite{DEMNUni_clustering_paper}:
\begin{equation}
P_{\mathrm{mm}}(k) \simeq 
    \begin{cases}
    P_{\mathrm{cc}}(k) & \text{for } k \ll k_{\mathrm{nr}} \\
    \left( 1 - f_{\nu} \right)^2 P_{\mathrm{cc}}(k) & \text{for } k \gg k_{\mathrm{nr}}.
    \end{cases} \label{eq:Pmm}
\end{equation}
The behaviour of $P_{\mathrm{mm}}(k)$ on intermediate scales is more complicated, but Eq.~(\ref{eq:Pmm}) shows the dominant influence that neutrinos have on the power spectrum of matter perturbations.

\subsection{Power spectra of voids \label{subsec:powspec_theory}}

For our analysis we make use of the matter and halo auto-power spectra, as well as the matter-void and halo-void cross-power spectra. We treat halos as proxies for galaxies. The model for the cross-power spectrum between halos and voids from reference~\cite{hamaus_voidPowSpec} is split into two terms, the first one is the one-void ($1 \mathcal{V}$) or shot noise term:

\begin{equation}
\label{voidcrosspowspec1}
P_{\mathrm{vh}}^{\left( 1 \mathcal{V} \right)} = \frac{1}{\bar{n}_\void \, \bar{n}_\halo } \int \frac{\text{d} n_\void(r_\void)}{\text{d} r_\void} \,
N_\halo (r_\void) \, u_\void(k | r_\void) \, \text{d}r_\void \,.
\end{equation}
It only considers correlations between $N_\halo$ halos and the void center within a given void of radius $r_\void$, the $\bar{n}_{\mathrm{i}}$ are the corresponding number densities and $u_\void(k | r_\void)$ describes the void density profile in Fourier space with normalization $u_\void(k \rightarrow 0) = 1$ to ensure the correct large-scale limit. The second is the two-void ($2 \mathcal{V}$) term that is responsible for the correlations between halos and void centers in distinct voids:
\begin{equation}
\label{voidcrosspowspec2}
P_{\mathrm{vh}}^{\left( 2 \mathcal{V} \right)} = \frac{1}{\bar{n}_\void   \bar{n}_\halo } \iint \frac{\text{d} n_\void(r_\void)}{\text{d} r_\void}
\, \frac{\text{d} n_\halo(m_\halo)}{\text{d} m_\halo} \,\, b_\void(r_\void) \,\, b_\halo(m_\halo)
\, u_\void(k | r_\void) \, P_{\mathrm{mm}}(k) \, \text{d}r_\void \, \text{d}m_\halo \,.
\end{equation}
Here, $n_\void(r_\void)$ is the number density function of voids of radius $r_\void$ and $n_\halo(m_\halo)$ the number density function of halos of mass $m_\halo$, with corresponding bias parameters $b_\void$ and $b_\halo$, neglecting terms of higher order in the bias. For a narrow range in void radii, the void-halo cross-power spectrum and the auto-power spectra of voids and halos simplify to \cite{hamaus_voidPowSpec}:

\begin{equation}
\label{void_and_galaxy_powspecs}
\begin{split}
P_{\mathrm{vh}}(k) & \simeq \, b_\void \, b_\halo \, u_\void (k) \,P_{\mathrm{mm}}(k) + \bar{n}^{-1}_\void \, u_\void(k) \,, \\
P_{\mathrm{vv}}(k) & \simeq \, b_{\void}^2 \, u_{\void}^2(k) \, P_{\mathrm{mm}}(k) + \bar{n}^{-1}_\void \,, \\
P_{\mathrm{hh}}(k) & \simeq \, b_{\halo}^2 \, P_{\mathrm{mm}}(k) + \bar{n}^{-1}_\halo \,.
\end{split}
\end{equation}

These equations show the direct relation between the different power spectra and the bias parameters. The inverse number densities describe the Poisson shot noise, which is modulated by $u_\void(k)$ for void-halo cross-correlations. However, in our analysis we will \CK{neglect} the shot noise from $P_{\mathrm{vh}}$, since it is much lower than the Poisson expectation due to the mutual exclusion of voids and halos~\cite[][]{hamaus_voidPowSpec,Chan2014,Platen2008}. Similar exclusion effects matter for correlations among massive halos~\cite{Hamaus2010,Baldauf2013}, but the overall amplitude is weaker and we neglect it here. \CC{Furthermore, we will focus our analysis on the large-scale clustering regime, where the void density profile reaches its large-scale limit, $u_\void(k)\rightarrow1$.}

\section{Data}
\label{sec:data}

\subsection{DEMNUni simulations \label{subsec:DEMNUNi_sims}}

\begin{figure}[!t]
\centering
\begin{subfigure}{1.0\textwidth}
  \centering
  \includegraphics[width=.99\linewidth,trim=0 10 0 0]{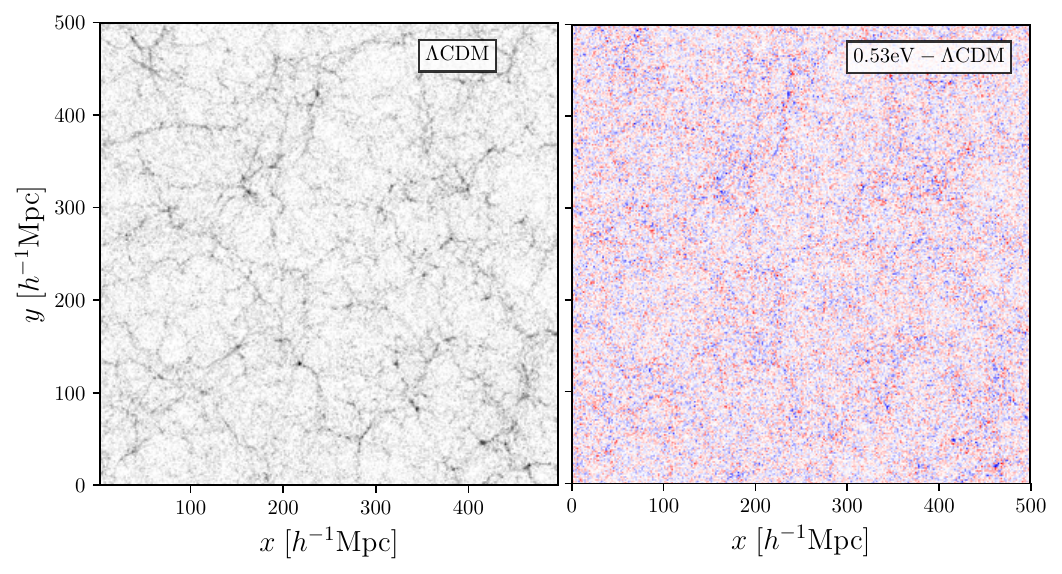}
\end{subfigure}%
\caption{Projected density field of CDM in a $ 50 \hMpc $ slice of the $\Lambda$CDM simulation (left) and differences in the projected density fields (right) between the $M_\nu=0.53$eV and the $\Lambda$CDM case (positive values in red, negative values in blue).}
\label{fig_simulation_sliceDM}
\begin{subfigure}{1.0\textwidth}
  \centering
  \includegraphics[width=.99\linewidth,trim=0 10 0 0]{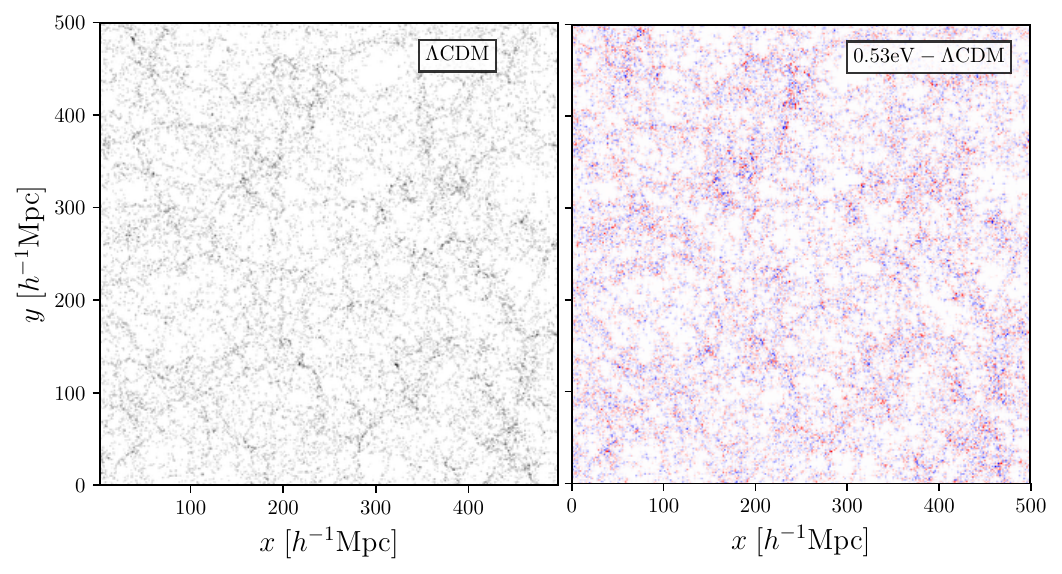}
\end{subfigure}%
\caption{Same as figure~\ref{fig_simulation_sliceDM}, but for halos instead of CDM.}
\label{fig_simulation_sliceHalo}
\end{figure}

This work uses snapshots from simulations of the ``Dark Energy and Massive Neutrino Universe'' (\texttt{DEMNUni} \cite{DEMNUni_clustering_paper,DEMNUni_weak_lensing_paper}) project at redshift $z = 0$. These simulations have been performed with the tree particle mesh-smoothed particle hydrodynamics (TreePM-SPH) code \textsc{gadget3} \cite{gadget2001}, with modifications to account for the presence of massive neutrinos implemented in reference \cite{Viel2010}. The \texttt{DEMNUni} simulations are characterized by their large volume, namely a box of side length $ 2 \, \hGpc$, following the evolution of $2048^3$ CDM and $2048^3$ neutrino particles when $\sumnu \neq 0$, allowing us to study the effects of massive neutrinos on large scales. The simulations are distinguished by their value of the sum of neutrino masses $\sumnu \equiv M_\nu$.

The relative energy densities of cold dark matter $\Omega_\mathrm{cdm}$ and neutrinos $\Omega_{\nu}$ vary for each different simulation with $\Omega_\mathrm{cdm} = 0.2700, 0.2659, 0.2628$ and $0.2573$ for the values $ M_\nu = 0.0, 0.17, 0.30$ and $0.53$eV, respectively. The other cosmological parameter values are chosen according to the $2013$ Planck results \cite{Planck2013Parameters}, with $\Omega_\mathrm{m} = 0.32$, $\Omega_\mathrm{b} = 0.05$, $n_s = 0.96$, $h = 0.67$ and $A_s = 2.1256 \times 10^{-9}$. Due to the fixed amplitude of primordial curvature perturbations $A_s$, the value of $\sigma_8$ changes with different $M_\nu$, where simulations with massive neutrinos have smaller values of $\sigma_8$ than the $\Lambda$CDM simulation with massless neutrinos. The mass of the simulated CDM particles therefore decreases with increasing values of $M_\nu$, due to the changes in $\Omega_\mathrm{cdm}$ while keeping $\Omega_\mathrm{m}$ and $\Omega_\mathrm{b}$ fixed. The halo catalogs have been produced via the friends-of-friends (FoF) and \textsc{subfind} algorithms that are included in \textsc{gadget3}~\cite[][]{gadget2001, Dolag2010}. The minimum number of member CDM particles has been set to 32, which corresponds to a minimum halo mass of $M_{\mathrm{min}} = 2.5 \times 10^{12} \hMsun$, resulting in a \CK{varying total number of halos for the different neutrino simulations}. In our analysis we did not use the full halo catalogs but instead fixed the halo number density, resulting in different minimum halo masses for each simulation, with all of them higher than the aforementioned $M_{\mathrm{min}}$.

\REF{Figures~\ref{fig_simulation_sliceDM} and \ref{fig_simulation_sliceHalo} present the projected density fields of both CDM and halos from a $50\hMpc$ slice of the $\Lambda$CDM simulation, as well as their differences between the $\Lambda$CDM and the neutrino simulation with $M_\nu=0.53$eV. As expected, the distribution of halos matches the one from the CDM particles. Voids in-between the visible clusters and filaments are apparent by eye. Differences in the voids between the $0.53$eV and the $\Lambda$CDM case are most apparent at adjoining lines of red and blue, indicating the void boundaries.
}

\subsection{Void finding \label{subsec:void_finding}}

The voids in this work have been identified with help of the Void IDentification and Examination toolkit {\textsc{vide}}\footnote{\url{http:www.cosmicvoids.net}} \cite{vide_general_paper}, which implements an enhanced version of \textsc{zobov} \cite{zobov-paper} to construct voids with a watershed algorithm. In the process of void finding \textsc{vide} performs a Voronoi tessellation of a given tracer distribution to define a density field, and identifies basins around local minima therein. Using the watershed transform \cite{watershed-2007}, \textsc{vide} then merges these basins and constructs a hierarchy of voids. In addition, \textsc{vide} imposes a density-based threshold within \textsc{zobov}, merging zones only if the density of their shared wall is below $20\%$ of the mean particle density $\bar{n}$. This prevents voids from extending deeply into overdense structures and limits the depth of the void hierarchy \cite{vide_general_paper}.

The result is a catalog of non-spherical voids. The volume $V$ of each void is calculated as the total volume of all Voronoi cells it is composed of. The effective radius of a void is defined as
\begin{equation}
    r_\void \equiv \left( \frac{3}{4 \pi} V \right)^{1/3} .
\end{equation}
Another important void property is the core density $\coreDens$. It is the density of the void's core particle $n_{\mathrm{core}}$, i.e. the density of the largest Voronoi cell of a void, expressed in units of the mean tracer density $\bar{n}_\tracer$,
\begin{equation}
    \coreDens \equiv \frac{n_{\mathrm{core}}}{\bar{n}_\tracer}.
\end{equation}
Lastly we define a property which we denote as the void \emph{compensation} $\Delta_\tracer$, where the index `t' (tracers) stands either for `h' (halos) or `m' (CDM) serving as tracers for the void finding with \textsc{vide}. This property is determined by the number of member particles $N_\tracer$ of a particular void and its volume,
\begin{equation}
\label{def_delta_N}
\Delta_\tracer \equiv \frac{N_\tracer}{\bar{n}_\tracer V} - 1.
\end{equation}
Thus, the compensation of a void specifies whether it contains more or less member particles than an average volume in the Universe of the same size. \CK{}\AP{Such voids are referred to as being \emph{over-} or \emph{undercompensated}, respectively~\cite{Hamaus2014b}. $\Delta_\tracer$ strongly correlates with the void bias $b_\void$: both quantities are positive for overcompensated, and negative for undercompensated voids. Exactly compensated voids with $N_\tracer=\bar{n}_\tracer V$ have $b_\void\simeq0$~\cite{Hamaus2014b}}.

Since \textsc{vide} only relies on the positions of tracer particles to find voids, both the distribution of CDM particles and the distribution of halos can be used for this. The halo catalogs of the \texttt{DEMNUni} simulations contain between $1.54 \times 10^7$ ($M_\nu = 0.0$eV) and $1.49 \times 10^7$ ($M_\nu = 0.53$eV) halos. We pick the $1.40 \times 10^7$ most massive halos of each simulation to perform our analysis with a fixed number density of $\bar{n}_{\mathrm{t}} \simeq 1.75 \times 10^{-3} \,(\hMpc)^3$. \EM{}\CK{}\AP{We choose to fix the halo number density instead of performing a constant mass cut to eliminate the impact of a varying tracer sparsity on void finding across simulations with different neutrino mass. This implies that the halo-mass ranges and hence the halo bias changes with $M_\nu$, which also impacts the properties of voids. However, even at fixed halo mass neutrinos affect the halo bias~\cite{LoVerde2014,Chiang2019}, so it is not possible to completely eliminate the neutrino dependence of halo bias for the identification process of voids. We therefore regard this halo bias dependence as a genuine impact that neutrinos have on voids that are identified in the distribution of halos with a given density. Note that in practice it is possible to achieve a given comoving tracer density, e.g., by constructing volume-limited samples from a redshift survey.}

Due to the fact that the process of finding voids is computationally expensive, we sub-sample the CDM particles to around $0.163 \%$ of the original particle number when finding voids in the distribution of CDM. This degree of sub-sampling is not chosen at random, but it results in the same tracer density as the one we fix for halos. \AP{It also guarantees our CDM- and halo-defined voids to have similar ranges in size} \REF{and eliminates possible differences that are merely due to Poisson noise in the tracer distribution, which is difficult to model or disentangle from other effects (see~\cite{vandeWeygaert2009} for an extensive study of tessellation based density and velocity field estimates)}. From here on, the voids we identify in the distribution of halos will be referred to as halo voids, whereas the voids found in the CDM distribution will be referred to as CDM voids. We do not attempt to identify voids in the total matter distribution (including neutrinos) here, as this adds a further complication to the void finding. However, voids in the 3D matter distribution are not directly accessible to observations anyway, so we only study CDM voids as a reference case to guide our physical intuition.

\section{Analysis \label{sec:analysis}}

For our analysis we focus on voids found in the distribution of CDM particles and halos. The presence of massive neutrinos affects both the number, as well as the clustering properties of voids~\CK{\cite{Massara2015,Kreisch2018}}. We use Eq.~(\ref{void_and_galaxy_powspecs}) to estimate the void bias $b_\void$ and the relative bias between voids and halos $b_\void / b_\halo$, respectively. In the case where the full matter distribution composed of CDM particles and neutrinos is available, we can use
\begin{equation}
\label{eq:void_bias_CDM}
 b_\void (k) = \frac{P_{\mathrm{vm}} (k)}{P_{\mathrm{mm}}(k)}
\end{equation}
to directly calculate the void bias. We have also considered using only the CDM component to define the void bias in Eq.~(\ref{eq:void_bias_CDM}), replacing the index `m' by the index `c', but we verified that our results are marginally affected by this choice. \CK{However, without knowledge of the matter distribution we can only determine the relative bias between voids and halos via}
\begin{equation}
\label{eq:void_bias_halos}
\frac{b_\void}{b_\halo} (k) =  \frac{P_{\mathrm{vh}} (k)}{P_{\mathrm{hh}}(k) - 1/ \Bar{n}_\halo}
\end{equation}
from the observable power spectra (see references~\cite{Pollina2017,Pollina2018} for corresponding quantities in configuration space). Note that for the matter field there is no shot noise in the power spectrum. The voids are divided into different bins in their core density $\coreDens$ and compensation $\Delta_\tracer$, the analysis is performed for each bin separately. We \CK{calculate the average large-scale bias from a range of low-$k$ values} and visually summarize our results in figures. Further, we perform the analysis for CDM voids and halo voids individually. For the case of halo voids, we additionally consider the total matter distribution (CDM and neutrinos combined) in the calculation of power spectra. The latter method more closely relates to observations using galaxies to identify voids, and weak lensing to probe the matter density field around them.

\subsection{Voids in cold dark matter \label{subsec:Voids_cdm}}

\begin{figure}[!t]
\centering
\begin{subfigure}{.5\textwidth}
  \centering
  \includegraphics[width=.99\linewidth, trim = 0 15 0 0]{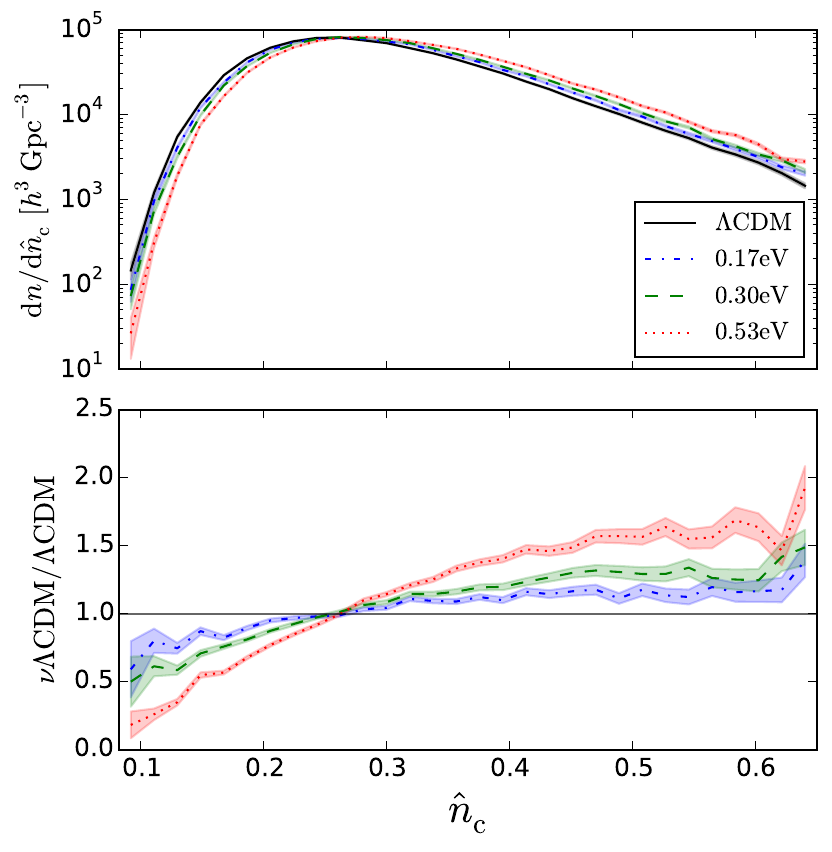}
\end{subfigure}%
\begin{subfigure}{.5\textwidth}
  \centering
  \includegraphics[width=.99\linewidth, trim = 8 15 0 0]{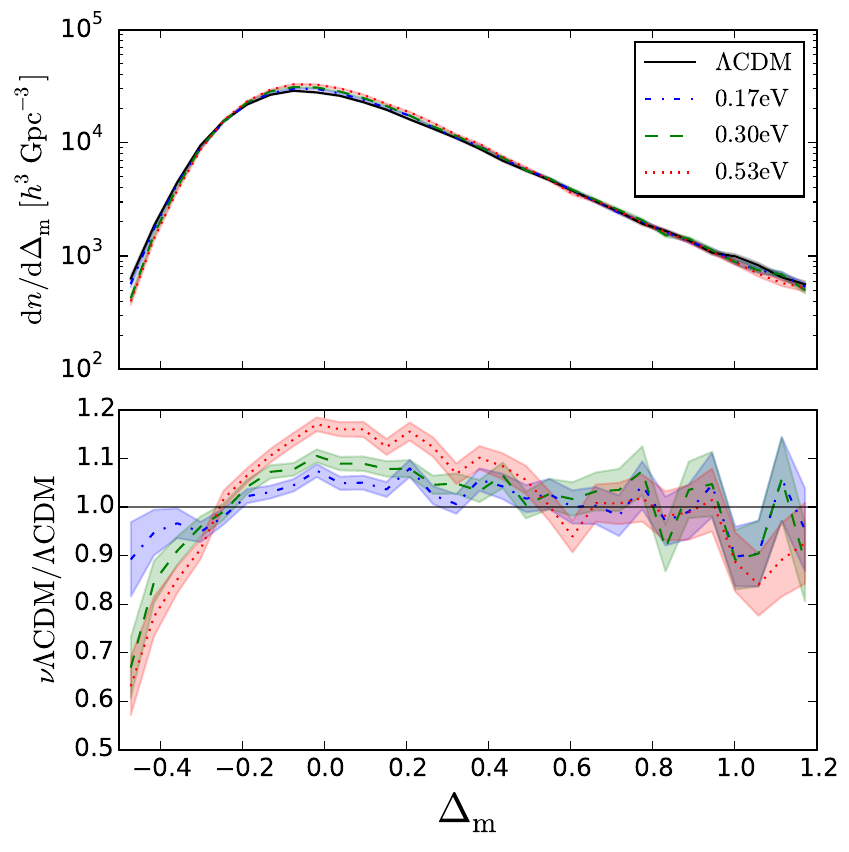}
\end{subfigure}
\caption{Top: Number density of CDM voids as a function of core density (left) and compensation (right) in different neutrino cosmologies. Bottom: Ratios of number densities in different neutrino cosmologies with respect to $\Lambda$CDM. Error bars assume Poisson statistics.}
\label{fig_void_density_CDM}
\end{figure}

\begin{figure}[!ht]
\centering
\resizebox{\hsize}{!}{
\includegraphics[width=\hsize,trim = 0 8 0 5,clip]{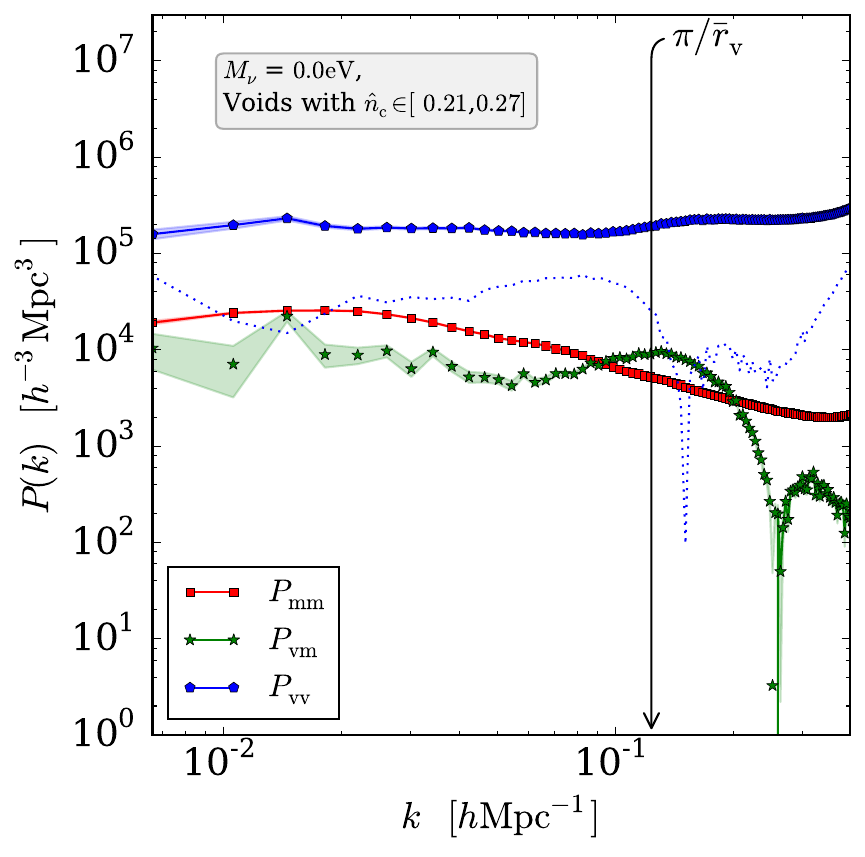}
\includegraphics[width=0.83\hsize,trim = 70 8 0 0,clip]{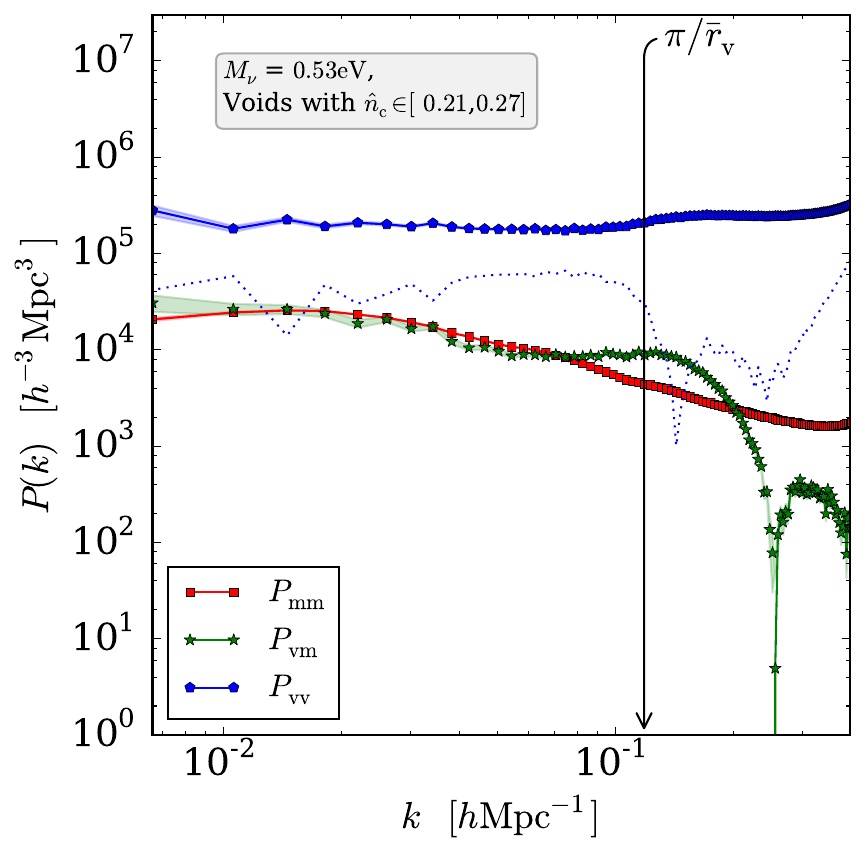}
}
\caption{Auto- and cross-power spectra of CDM voids with core density $\coreDens \in [0.21,0.27]$ for $M_\nu=0.0$eV (left) and $M_\nu=0.53$eV (right). The dotted blue line shows $P_{\mathrm{vv}}$ with Poisson shot noise subtracted. Not connected points have negative values ($P_{\mathrm{vm}}$ \CK{at $k\lesssim0.25\,h\mathrm{Mpc}^{-1}$}). Shaded areas show $1 \sigma$ uncertainties estimated from the scatter of modes in each $k$-bin. An arrow indicates the scale of the average void size (void-exclusion scale) in each cosmology.}
\label{fig_void_powspec_CDM}
\end{figure}

The total number of voids we find with \textsc{vide}, using CDM particles as tracers, lies in between $133\,349$ ($0.0$eV) and $145\,486$ ($0.53$eV), with an increasing number for increasing neutrino masses. \CK{It happens as large voids get fragmented into smaller ones due to the additional substructure induced by neutrinos inside voids~\cite{Massara2015,Kreisch2018}}. This is supported by \CC{the left panel of} figure~\ref{fig_void_density_CDM}, where voids' core densities tend to increase with increasing $M_\nu$, leaving fewer voids of low core density. \AP{As neutrinos freely stream deeply into voids they gravitationally attract more mass to remain there~\cite{Massara2015}}. The distribution of voids as a function of their compensation, \CC{as shown in the right panel of figure~\ref{fig_void_density_CDM}}, indicates that for higher $M_\nu$, \CK{}\AP{more voids exhibit values around $\Delta_\matter = 0$}, while their abundance at the far ends of the distribution is suppressed. \AP{Hence, in neutrino cosmologies more CDM voids tend to be compensated.}

\begin{figure}
\centering
\begin{minipage}{.5\textwidth}
  \centering
  \includegraphics[width=1.\linewidth, trim=0 6 0 8 ]{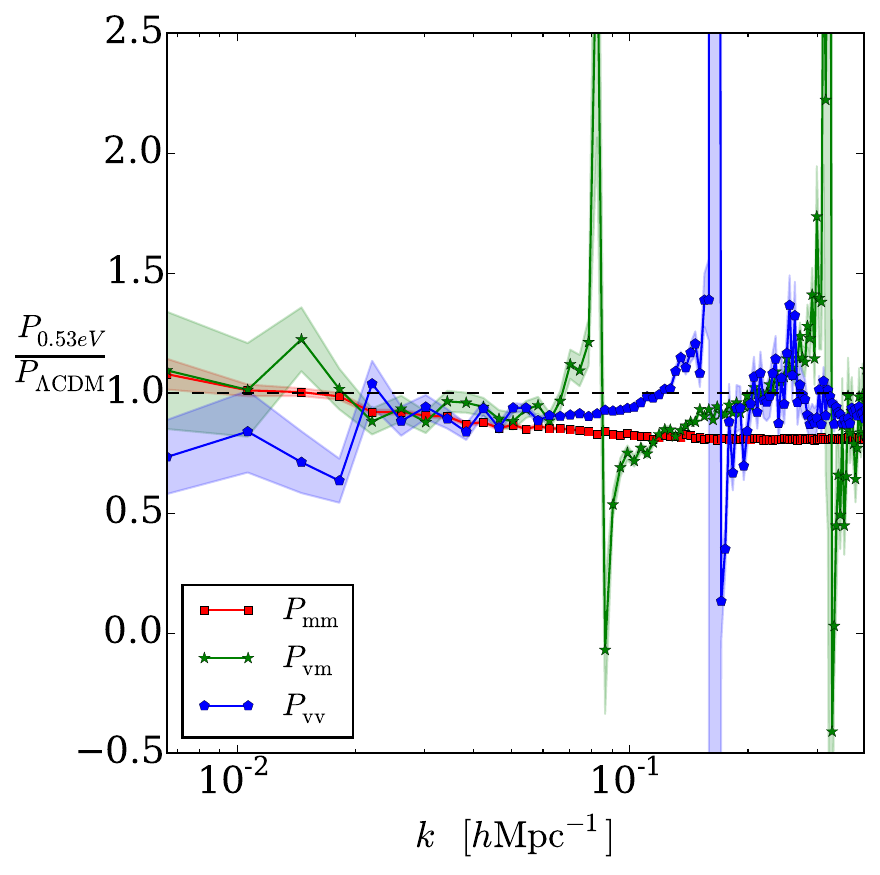}
  \captionsetup{width=.95\linewidth}
  \captionof{figure}{Ratio of power spectra (matter and void auto-, as well as void-matter cross-power) between a cosmology with massive neutrinos ($M_\nu=0.53$eV) and $\Lambda$CDM ($M_\nu=0$). Here we use the entire CDM void sample.}
  \label{fig_void_Divided_PowSpec_CDM}
\end{minipage}%
\begin{minipage}{.5\textwidth}
  \centering
  \includegraphics[width=.92\linewidth,trim=0 0 0 0 ]{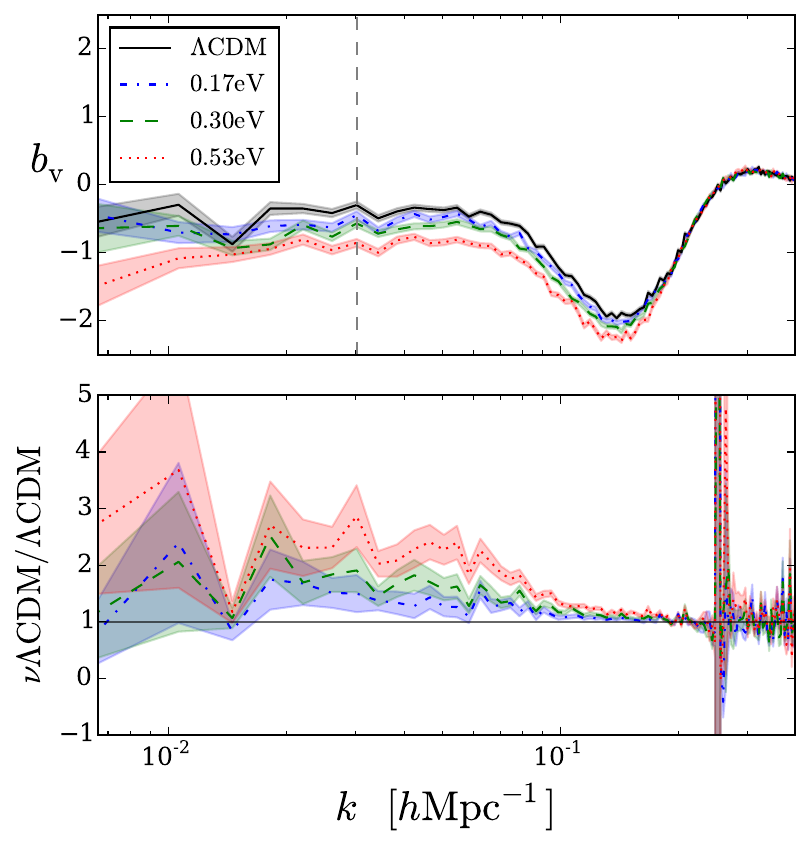}
  \captionsetup{width=.95\linewidth}
  \captionof{figure}{Void bias estimated via Eq.~(\ref{eq:void_bias_CDM}) for CDM voids with core density $\coreDens \in [0.21,0.27]$. The vertical line indicates the limit for large-scale averages. Ratios w.r.t. the $\Lambda$CDM case are shown in the bottom.}
  \label{fig_void_bias_CDM_coreDens021to027}
\end{minipage}
\end{figure}

Figure~\ref{fig_void_powspec_CDM} depicts the matter and void auto-power spectra, as well as the void-matter cross-power spectrum in a \AP{representative} bin with $\coreDens \in [0.21,0.27]$ for the $\Lambda$CDM simulation and for the $\nu \Lambda$CDM simulation with the highest mass of $0.53$eV. The matter power spectrum $P_{\mathrm{mm}}$ experiences an expected scale-dependent suppression due to neutrino free streaming~\CK{\cite{Lesgourgues2006}}. A similar trend is present in the cross-power spectrum $P_{\mathrm{vm}}$ between voids and matter, \CC{which inherits the scale-dependent suppression from $P_{\mathrm{mm}}$}. However, $P_{\mathrm{vm}}$ is negative on large scales due to the negative bias of the selected void sample, and only turns positive on small scales. Because $P_{\mathrm{vv}}$ is dominated by shot noise, we will not use it \CK{to calculate} the void bias in the following. \AP{However, we note the step-like feature around $k\simeq\pi/\bar{r}_\void$, which is caused by the mutual exclusion of voids due to their large extent~\cite{hamaus_voidPowSpec,Chan2014,Platen2008}, and that it is slightly shifted by the influence of neutrinos~\cite{Kreisch2018}. The small upturn at $k>0.3\,h\mathrm{Mpc}^{-1}$ is due to finite grid-size effects in the calculation of the power spectra}.

In order to showcase the relative effects more clearly, figure~\ref{fig_void_Divided_PowSpec_CDM} presents the ratio of the three power spectra between $\nu\Lambda$CDM and $\Lambda$CDM, in this case for \AP{the entire void sample} found in CDM. The large-scale power spectrum amplitude of the full void sample is less affected by neutrinos and has a positive bias, which allows us to discern the scale-dependent effects more clearly. \AP{Errors on the ratios of power spectra are calculated via Gaussian error propagation, assuming the power spectra are uncorrelated. Because the initial seeds of the simulations are identical, this is a conservative estimate and overestimates the true error bars in all quantities involving ratios of power spectra (e.g., the bias).} We notice the expected damping of the matter power spectrum caused by massive neutrinos \CC{towards smaller mildly nonlinear} scales, i.e. higher values of~$k$~\CK{\cite{Lesgourgues2006}}. \CK{Both void auto- and void-matter cross-power spectra exhibit a similar suppression as the matter power spectrum on mildly nonlinear scales, but a more complicated structure on small scales. In order to examine the origin of these effects, it is instructive to focus on the void bias $b_\void$ directly via Eq.~(\ref{eq:void_bias_CDM}).}

This is presented for the different neutrino cosmologies in figure~\ref{fig_void_bias_CDM_coreDens021to027}, using CDM voids in the core-density range of $\coreDens \in [0.21,0.27]$. Especially on large scales (\CK{left to the vertical dashed line at $k=0.03\,h\mathrm{Mpc}^{-1}$}) we notice clear differences between the simulations, with more negative values of $b_\void$ for higher masses $M_\nu$ in this void bin. In absolute value this resembles the behavior of halo bias, which increases with increasing neutrino masses~\cite{Kreisch2018,DEMNUni_clustering_paper,Marulli2011,LoVerde2014}. On smaller scales we observe a dip in the bias, as already found in reference~\cite{hamaus_voidPowSpec}, with again noticeable differences between the different neutrino masses. These are not exclusively due to a general scaling in the amplitude of the void bias, but also depend on wavenumber $k$ \CK{(consistent with what was found using void auto-power spectra in~\cite{Kreisch2018})}. \REF{Other small-scale neutrino effects have already been analyzed in~\cite{Massara2015} via void density profiles in configuration space. In this paper we focus on the large-scale linear bias, which already exhibits a rich phenomenology that can be probed with current and future surveys, and leave the analysis of the small-scale regime in Fourier space for future work.}

\begin{figure}[!t]
\centering
\begin{subfigure}{.5\textwidth}
  \centering
  \includegraphics[width=.99\linewidth]{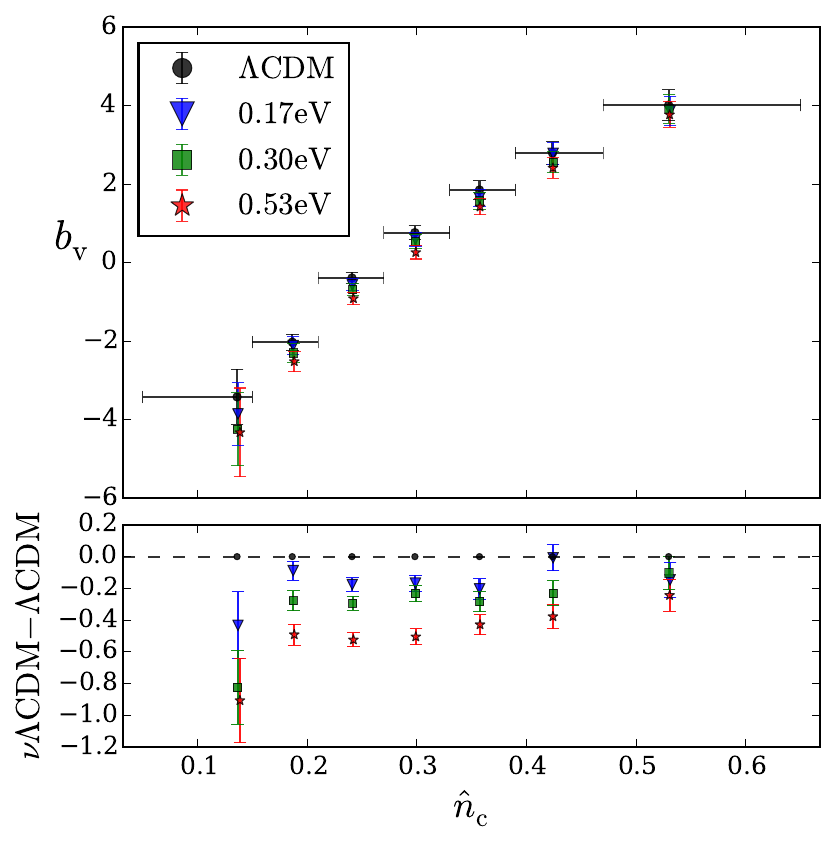}
  \label{fig_CDM_coreDens_summary}
\end{subfigure}%
\begin{subfigure}{.5\textwidth}
  \centering
  \includegraphics[width=.99\linewidth]{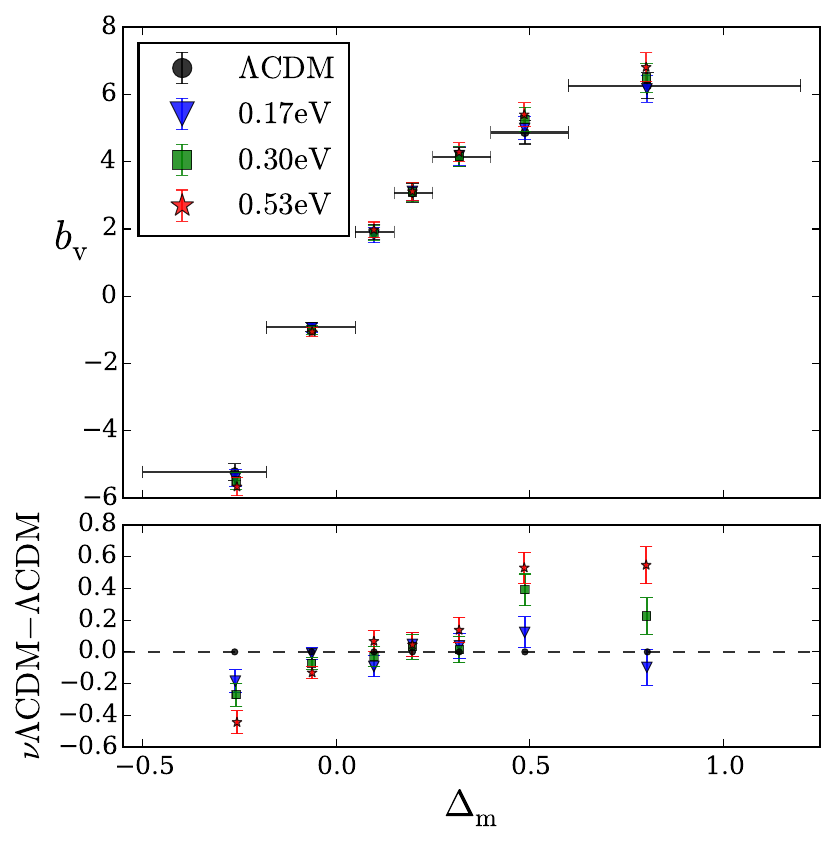}
  \label{fig_CDM_deltaN_summary}
\end{subfigure}
\vspace{-15pt}
\caption{Large-scale void bias (top) and its difference (bottom) between $M_\nu \neq 0$ ($\nu \Lambda$CDM) and $M_\nu = 0$ ($ \Lambda$CDM) cosmologies in bins of core density (left) and compensation (right) for CDM voids. \CK{Vertical error bars assume uncorrelated power spectra and are therefore maximally conservative}. In the top panels they are additionally enlarged by a factor of $5$ for visibility. Horizontal error bars indicate the size of each bin (same for all 4 simulations), and data points are located at the mean value of each bin.}
\label{fig_CDM_summary}
\end{figure}

Figure~\ref{fig_CDM_summary} shows the average large-scale values of $b_\void$ for all bins in $\coreDens$ and $\Delta_\matter$, \CC{where} \CK{vertical error bars in the top panels have been enlarged by a factor of 5} for better visibility. The averaging is performed for scales between $k_{\mathrm{min}} = 0.007\,h\mathrm{Mpc}^{-1}$ and $k_{\mathrm{max}} = 0.030\,h\mathrm{Mpc}^{-1}$, \CK{as indicated by the vertical dashed line in figure~\ref{fig_void_bias_CDM_coreDens021to027}}. We observe an overall trend that the void bias increases with increasing core density and increasing void compensation. We further note that massive neutrino simulations have distinct $b_\void$ values compared to the $\Lambda$CDM counterpart, and these differences follow the intuitive trend of increasing for larger values of $M_\nu$~\CK{(cf.~\cite{Massara2015,Kreisch2018})}. For the bins in core density we see that the average void bias \CC{becomes more negative overall, but most strongly at negative bias values}. For voids sorted by their compensation, the values of $b_\void$ also clearly differentiate between the changing sum of neutrino masses in most $\Delta_\matter$ bins. Here, the amplitude of $b_\void$ experiences a boost for $M_\nu \neq 0$, increasing \CC{the absolute value of void bias} compared to the $\Lambda$CDM values. \CK{Intuitively, this is in accord with the known trend of an increased number of small CDM voids in neutrino cosmologies~\cite{Massara2015,Kreisch2018}: initial density fluctuations of a given size and Lagrangian bias will form smaller voids at late times. Hence, these voids will have the Eulerian bias of larger voids in $\Lambda$CDM, which typically have a more negative bias in the regime $\Delta_\matter<0$~\cite{hamaus_voidPowSpec}. However, this trend seems to be reversed for overcompensated voids with $\Delta_\matter>0$ according to the right panel of figure~\ref{fig_CDM_summary}, \CC{a trend that looks more similar to what happens for halos}, which argues for another mechanism to be at play here. We defer a more detailed analysis of this to future work.}

\subsection{Voids in halos \label{subsec:Voids_halos}}

\begin{figure}[!t]
\centering
\begin{subfigure}{.5\textwidth}
  \centering
  \includegraphics[width=.99\linewidth]{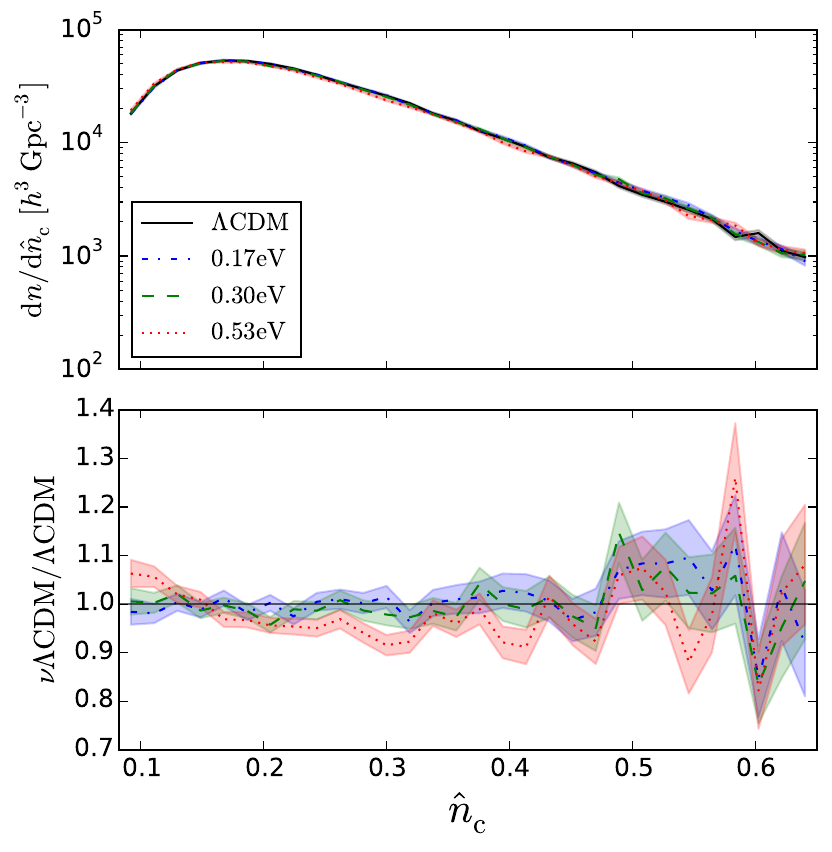}
\end{subfigure}%
\begin{subfigure}{.5\textwidth}
  \centering
  \includegraphics[width=.99\linewidth, trim = 0 0 0 0]{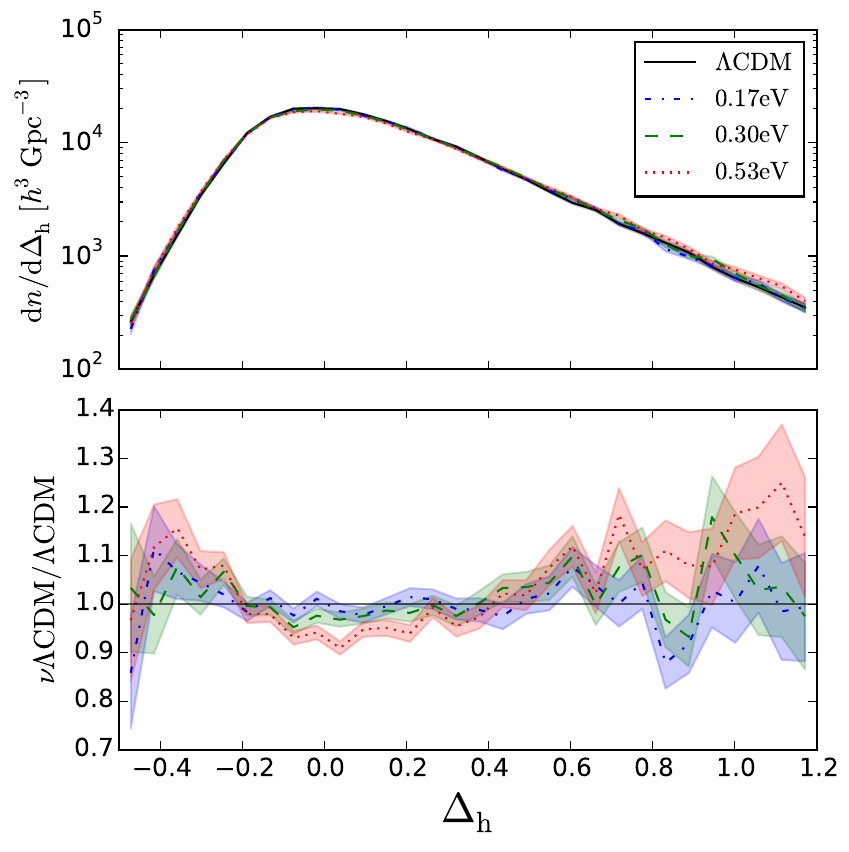}
\end{subfigure}
\vspace{-0.2cm}
\caption{Same as figure \ref{fig_void_density_CDM} for halo voids.}
\label{fig_void_density_halos}
\end{figure}

When we use the halo distribution for void finding we end up with $93\,214$ ($0.0$eV) and $91\,128$ ($0.53$eV) voids, respectively. Hence, now the number of voids decreases with increasing $M_\nu$, in contrast to the CDM case above. \CK{}\AP{The same inversion has already been observed in reference~\cite{Kreisch2018}, which links it to the halo-bias dependence on massive neutrinos}. \REF{For a more in-depth analysis of the halo mass function and its dependence on massive neutrinos we refer to~\cite{DEMNUni_clustering_paper}.} The number density distribution of halo voids depending on $\coreDens$ and $\Delta_\halo$ is shown in figure~\ref{fig_void_density_halos}. Compared to the distribution of CDM voids in figure~\ref{fig_void_density_CDM} we now find much smaller differences. In the distribution as a function of core density we do not observe the shift to voids with higher values of $\coreDens$ in $M_\nu \neq 0$ simulations, most noticeable is a slight increase in void numbers for \CC{very} small core densities with a slight decrease for larger values. Similarly, in the distribution of void compensations we observe an opposite effect from the case of CDM voids: now the number of voids in massive neutrino simulations tends to be suppressed for intermediate values of $\Delta_\halo$ around zero, but instead exhibits more voids at small and large~$\Delta_\halo$.


\begin{figure}[!h]
\centering
\begin{subfigure}{.5\textwidth}
  \centering
  \includegraphics[width=.99\linewidth]{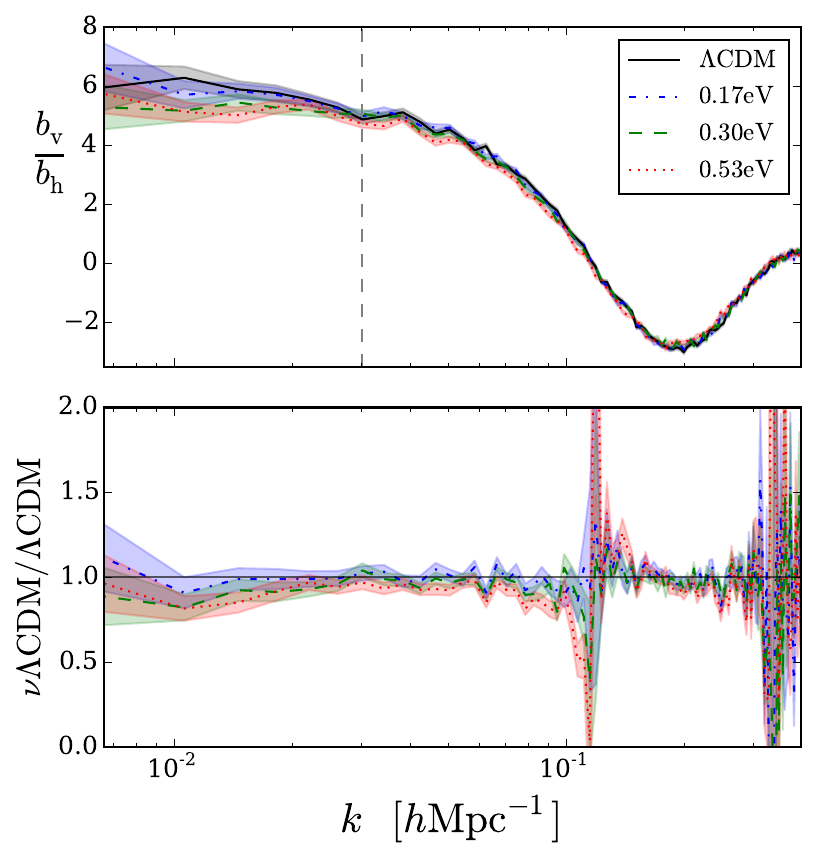}
\end{subfigure}%
\begin{subfigure}{.5\textwidth}
  \centering
  \includegraphics[width=.99\linewidth]{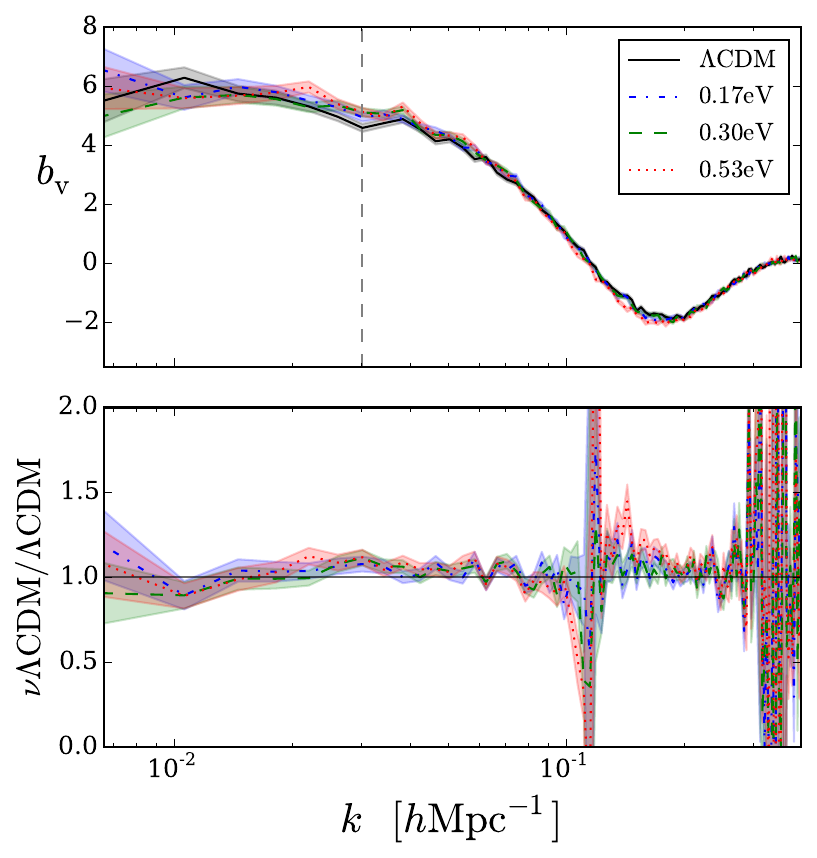}
\end{subfigure}
\caption{Relative bias for halo voids correlated with halos ($b_\void/b_\halo$, top left), and halo voids correlated with matter ($b_\void$, top right). Bottom panels show ratios of this bias from different neutrino cosmologies with respect to $\Lambda$CDM. All voids have compensation $\Delta_\halo \in [0.4,0.6]$. The vertical line indicates the limit for large-scale averages.}
\label{fig_void_bias_halos_deltaN015to025}
\end{figure}

\begin{figure}[!t]
\centering
\begin{subfigure}{.5\textwidth}
  \centering
  \includegraphics[width=.99\linewidth]{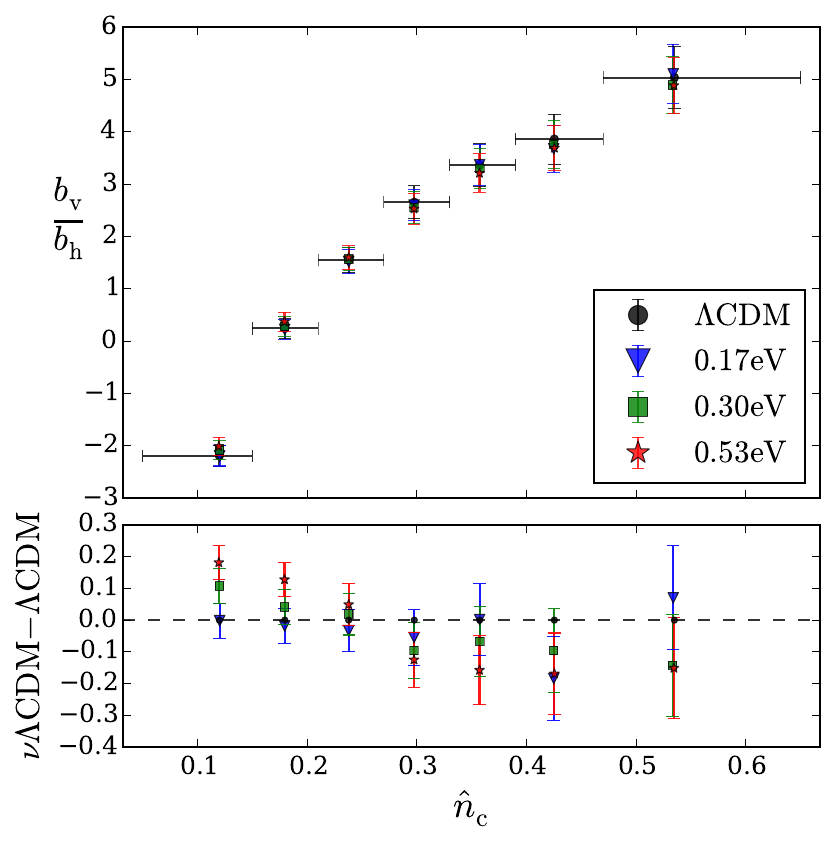}
\end{subfigure}%
\begin{subfigure}{.5\textwidth}
  \centering
  \includegraphics[width=.99\linewidth]{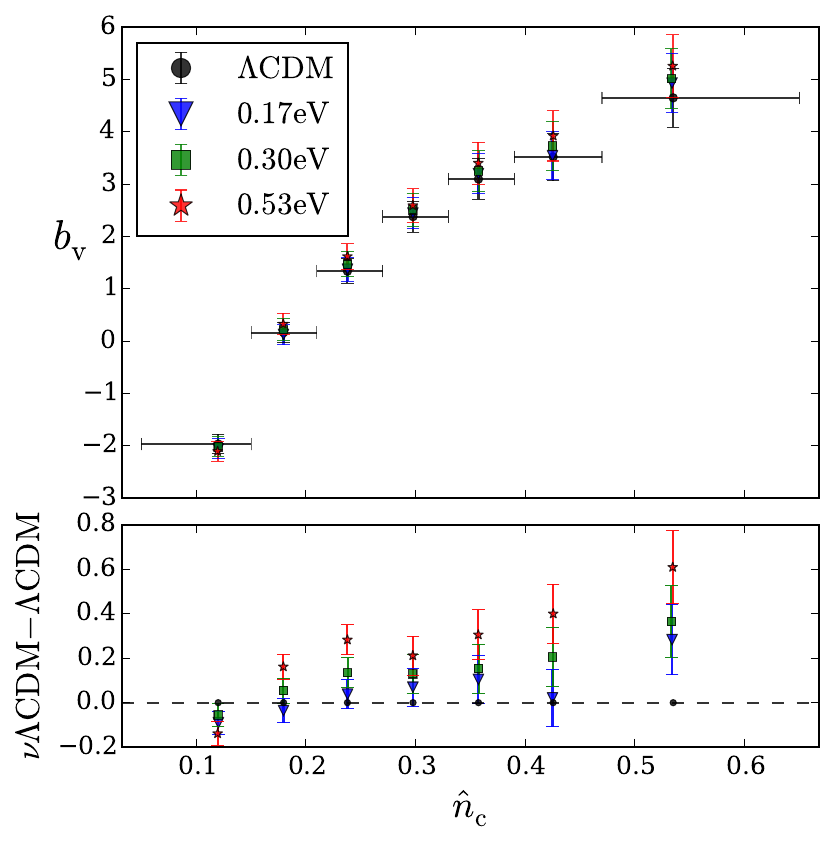}
\end{subfigure}
\caption{Large-scale relative bias for halo voids (top) and its difference (bottom) between $M_\nu \neq 0$ ($\nu \Lambda$CDM) and $M_\nu = 0$ ($\Lambda$CDM) in bins of core density. Halo voids are either correlated with halos (left), or matter (right).}
\label{fig_void_halos_coreDens_summary}
\end{figure}

\begin{figure}[!h]
\centering
\begin{subfigure}{.5\textwidth}
  \centering
  \includegraphics[width=.99\linewidth]{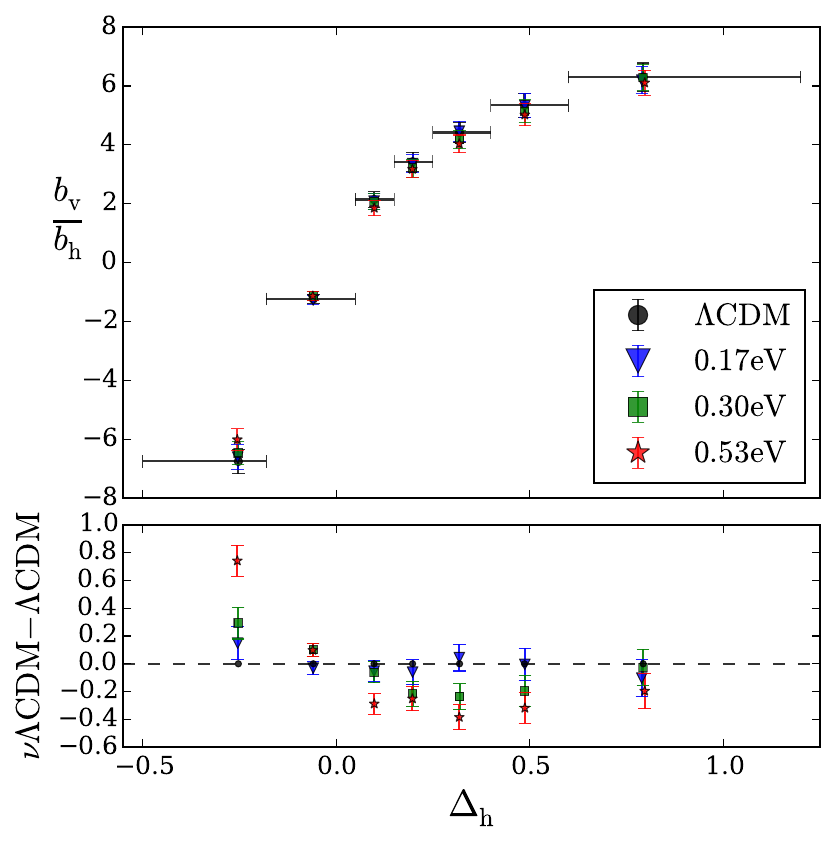}
\end{subfigure}%
\begin{subfigure}{.5\textwidth}
  \centering
  \includegraphics[width=.99\linewidth]{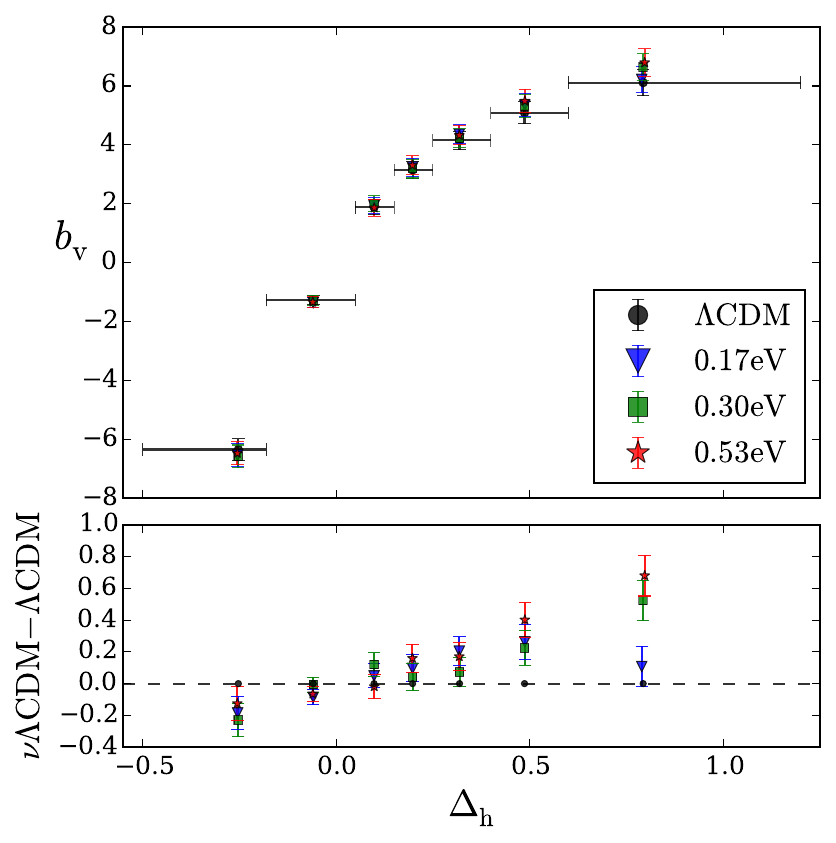}
\end{subfigure}
\caption{Same as figure \ref{fig_void_halos_coreDens_summary} for the bins in compensation $\Delta_\halo$.}
\label{fig_void_halos_deltaN_summary}
\end{figure}

Calculating the void bias for halo voids with $\Delta_\halo \in [0.4,0.6]$, once correlated with halos ($b_\void / b_\halo$ with Eq.~(\ref{eq:void_bias_halos})) and once correlated with the total matter ($b_\void$ with Eq.~(\ref{eq:void_bias_CDM})), yields the results shown in figure~\ref{fig_void_bias_halos_deltaN015to025}. We notice that including the total matter field (CDM and neutrinos) for calculating power spectra in the analysis of halo voids increases the sensitivity to signatures from massive neutrinos. Differences occur mostly on large scales and near the dip in the void bias at smaller scales, similar to the results from CDM voids, however with smaller amplitude \AP{and partly indistinguishable within the error}. In figure~\ref{fig_void_halos_coreDens_summary} and figure~\ref{fig_void_halos_deltaN_summary} we present the summary of our large-scale bias analysis of voids found in halos, both as a function of core density and for different void compensations. When using only halos, the bins with $\Delta_\halo \in [-0.5,-0.18]$ and $\Delta_\halo \in [0.25,0.4]$ show the clearest differences in void bias, although some error bars still overlap. \CK{Moreover, there is a hint of an inverted trend compared to the CDM void case in the right panel of figure~\ref{fig_CDM_summary} (cf.~\cite{Kreisch2018})}. When correlating halo voids with matter \CK{to calculate $b_\void$ for halo voids via $P_{\void\matter}$ and $P_{\matter\matter}$ with Eq.~(\ref{eq:void_bias_CDM}) directly}, we observe more distinct signatures from neutrinos. Again the differences between $\nu \Lambda$CDM and $\Lambda$CDM tend to increase with higher neutrino masses. Note also that, like in the analysis for CDM voids, the amplitude of $b_\void$ increases when sorting voids by their values of $\Delta_\halo$, so the inversion seen in the left panel of figure~\ref{fig_void_halos_deltaN_summary} is revoked. However, we observe an increase of the value of $b_\void$ from massive neutrinos when we sort the voids by core density \CC{(left panel of figure~\ref{fig_void_halos_coreDens_summary})}, which is still opposite to the case of CDM voids above.

\section{Model\label{sec:discussion}}

\begin{figure}[!t]
\centering
\begin{subfigure}{.5\textwidth}
  \centering
  \includegraphics[width=.99\linewidth,trim=0 10 0 0]{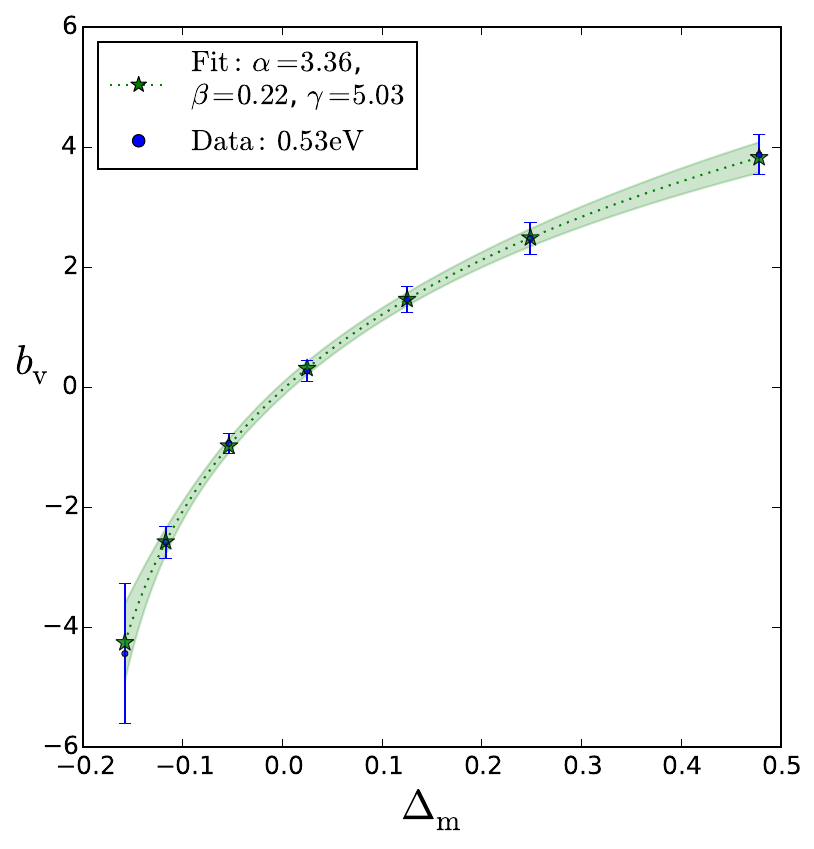}
\end{subfigure}%
\caption{Fit of Eq.~(\ref{eq:fit_function}) to the average large-scale void bias from the $M_\nu=0.53$eV simulation for CDM voids in each bin of core density $\coreDens$. The $x$-axis shows the fit in the mean values of compensation $\Delta_\matter$ that correspond to each bin of $\coreDens$. Errors in the data and the fit are enlarged by a factor of $5$ for visibility.}
\label{fig_single_fits_CDM_coreDens}
\end{figure}

\begin{figure}[!ht]
\centering
\begin{subfigure}{.5\textwidth}
  \centering
  \includegraphics[width=.99\linewidth]{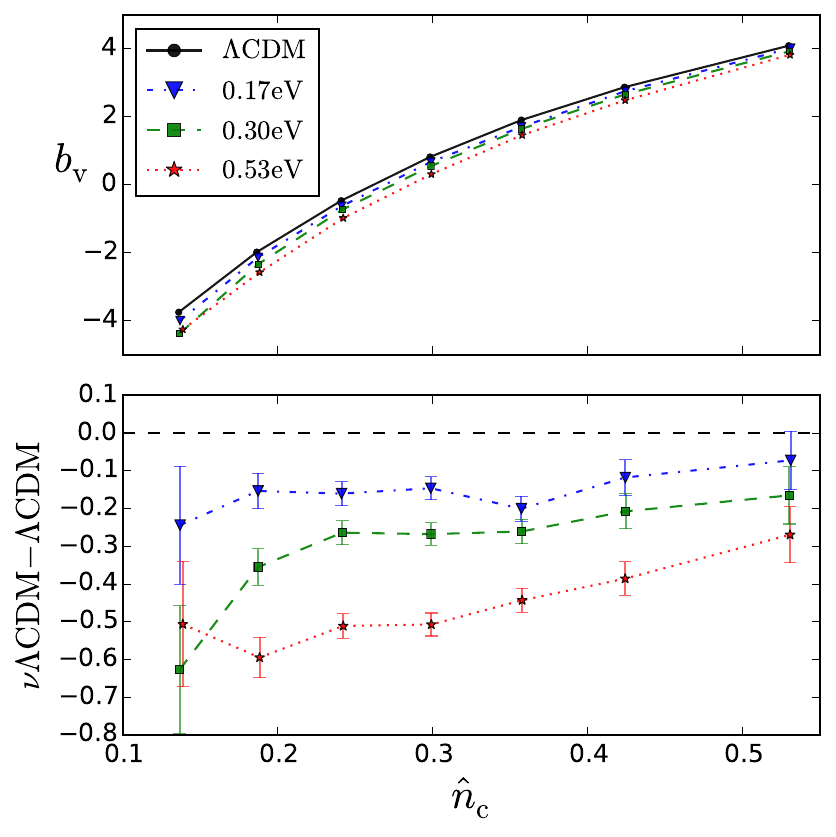}
\end{subfigure}%
\begin{subfigure}{.5\textwidth}
  \centering
  \includegraphics[width=.99\linewidth, trim = 0 0 10 10]{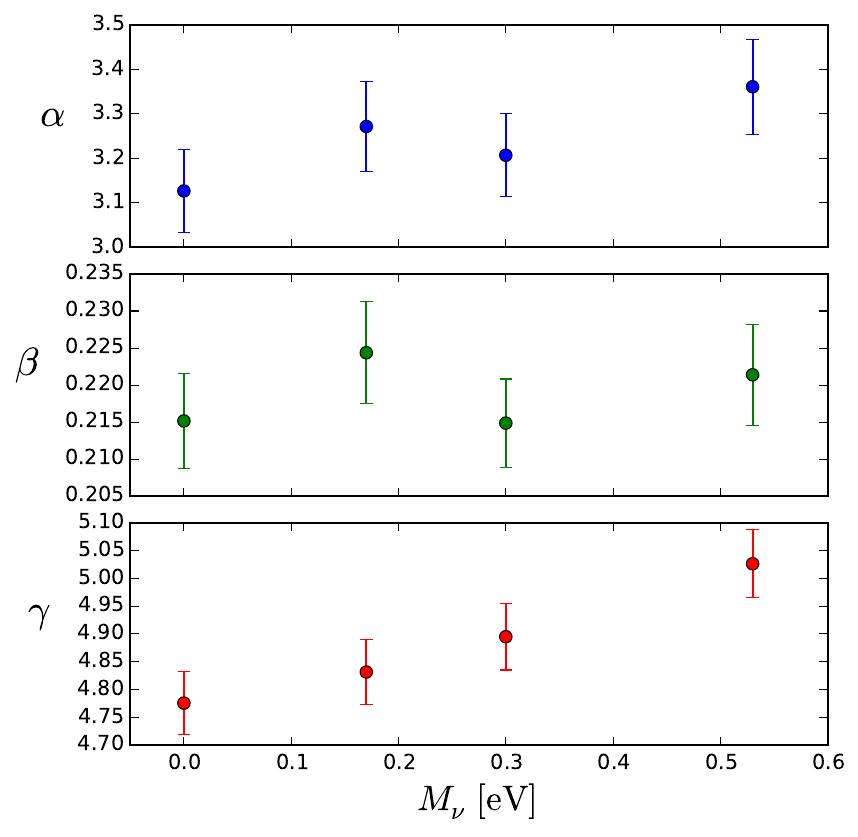}
\end{subfigure}
\caption{Left: Fit of Eq.~(\ref{eq:fit_function}) to all 4 simulations (top) in the core density bins for CDM voids, and differences between \AP{fits in} $\nu\Lambda$CDM and $\Lambda$CDM (bottom). Right: Corresponding best-fit parameter values.}
\label{fig_fit_summary_CDM_coreDens}
\end{figure}

\begin{figure}[!ht]
\centering
\begin{subfigure}{.5\textwidth}
  \centering
  \includegraphics[width=.99\linewidth]{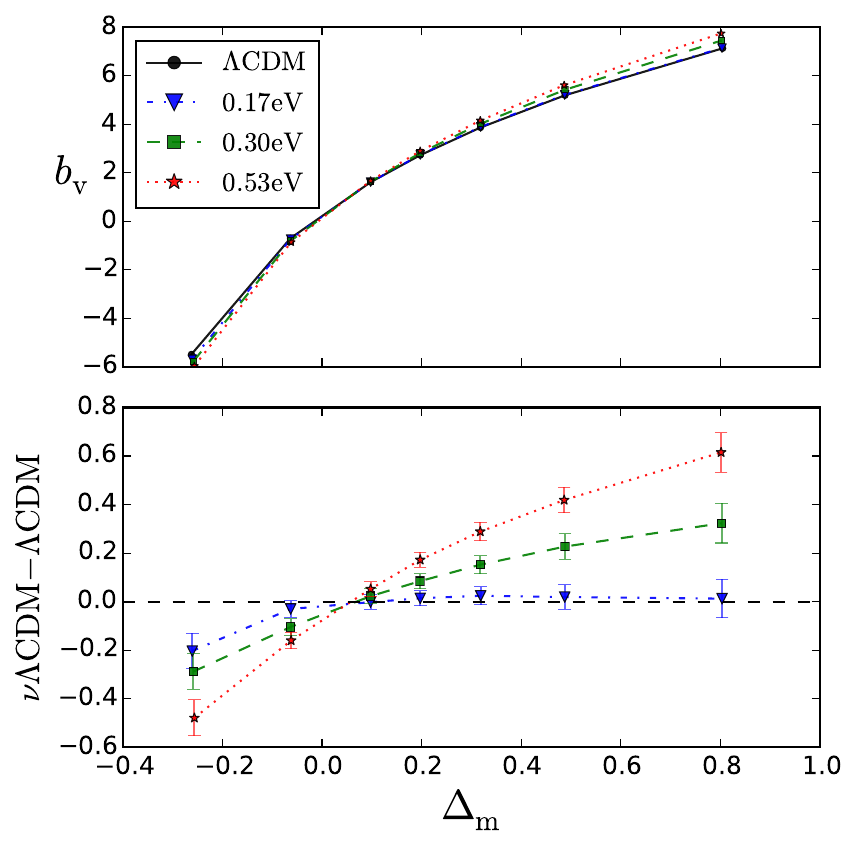}
\end{subfigure}%
\begin{subfigure}{.5\textwidth}
  \centering
  \includegraphics[width=.99\linewidth, trim = 3 6 0 10]{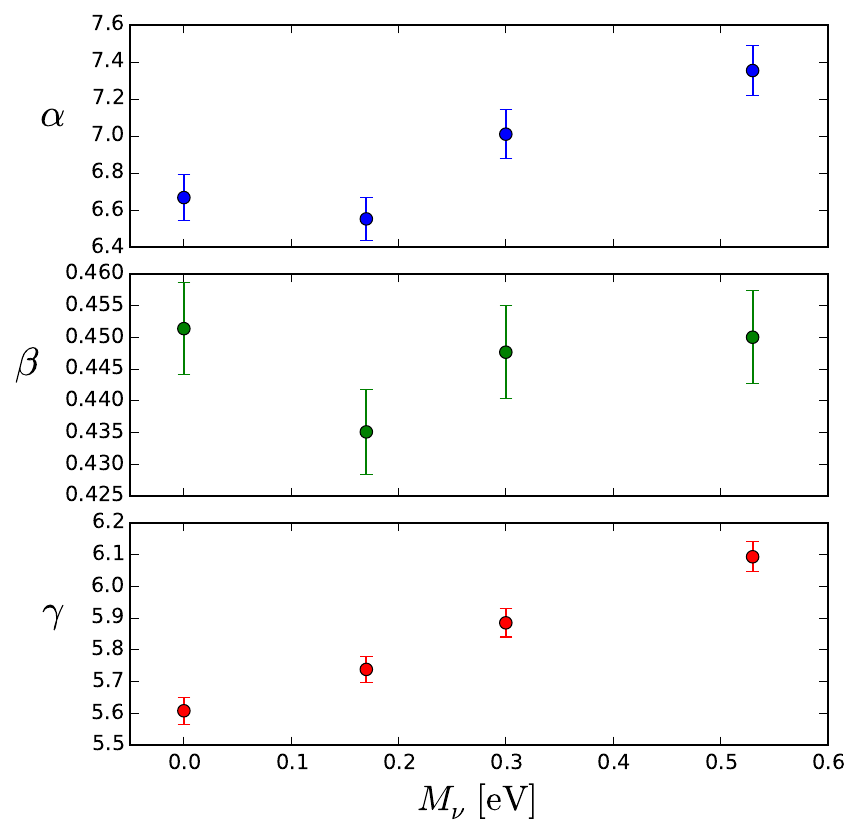}
\end{subfigure}
\caption{Same as figure \ref{fig_fit_summary_CDM_coreDens} for bins in compensation $\Delta_\matter$.}
\label{fig_fit_summary_CDM_deltaN}
\end{figure}

\begin{figure}[!ht]
\centering
\begin{subfigure}{.5\textwidth}
  \centering
  \includegraphics[width=.99\linewidth]{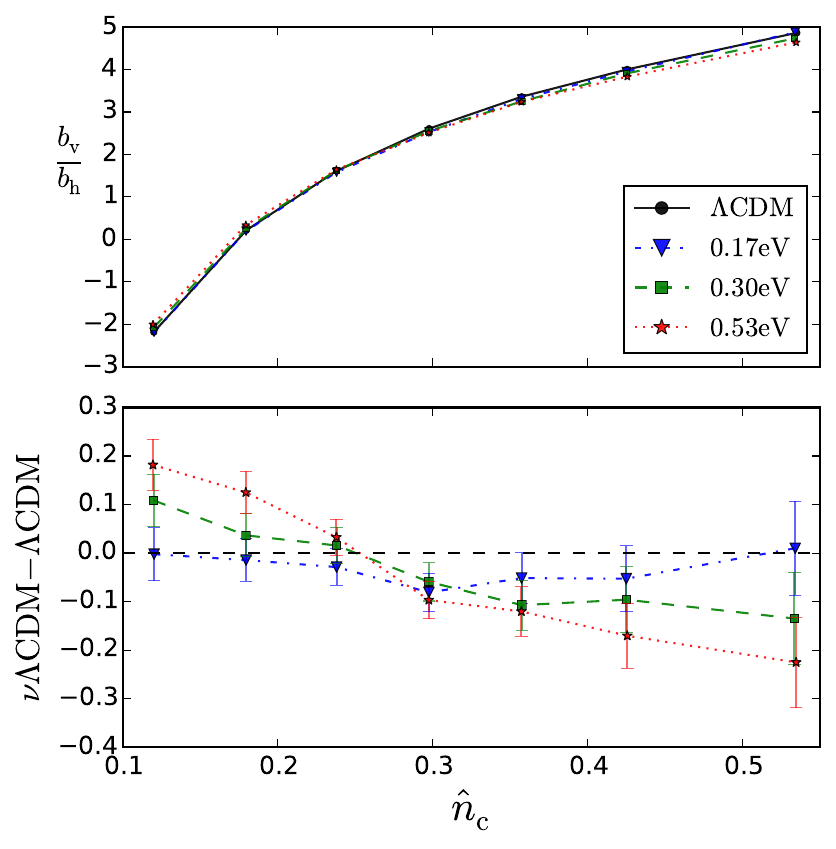}
\end{subfigure}%
\begin{subfigure}{.5\textwidth}
  \centering
  \includegraphics[width=.99\linewidth, trim = 10 5 0 10]{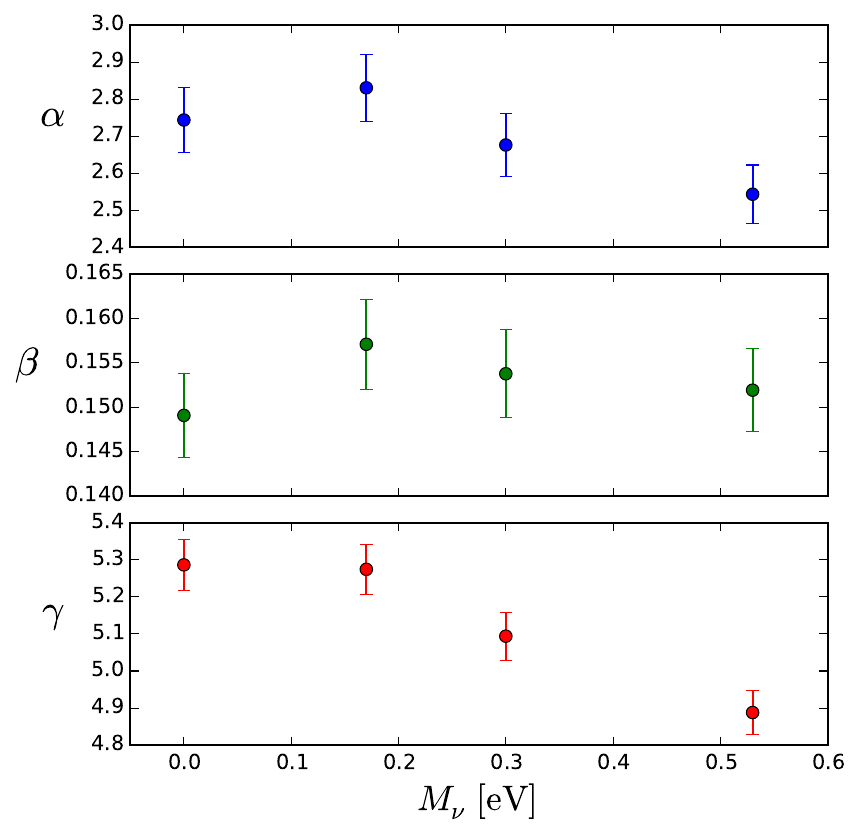}
\end{subfigure}
\caption{Same as figure \ref{fig_fit_summary_CDM_coreDens} for halo voids correlated with halos, bins in core density $\coreDens$.}
\label{fig_fit_summary_Halos_coreDens}
\end{figure}

\begin{figure}[!ht]
\centering
\begin{subfigure}{.5\textwidth}
  \centering
  \includegraphics[width=.99\linewidth]{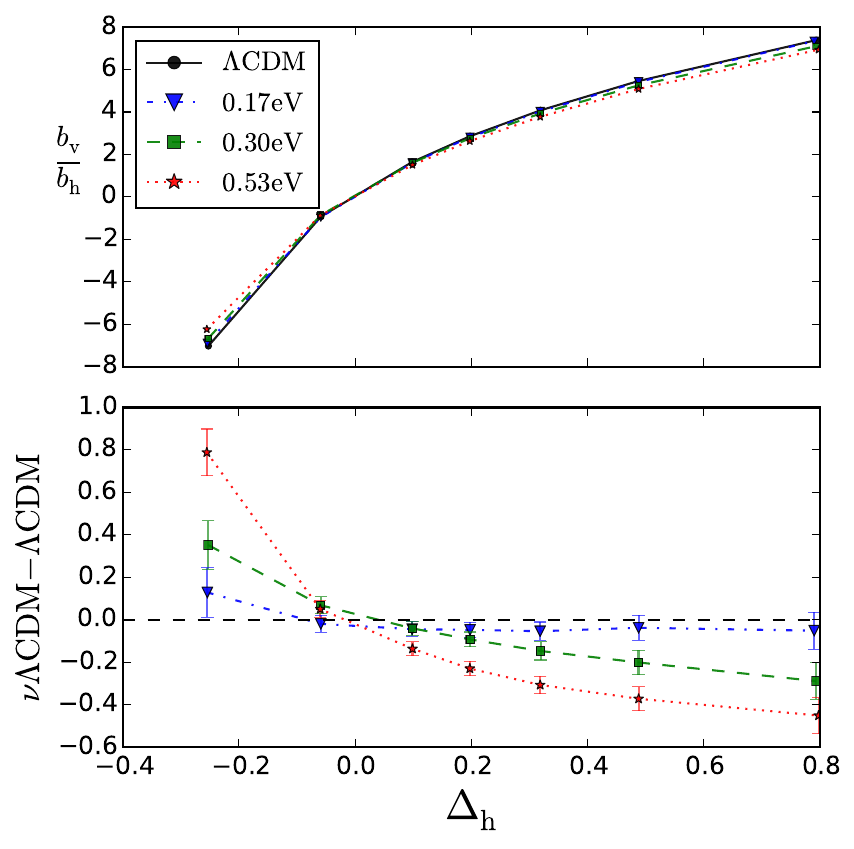}
\end{subfigure}%
\begin{subfigure}{.5\textwidth}
  \centering
  \includegraphics[width=.99\linewidth, trim = 4 8 0 10]{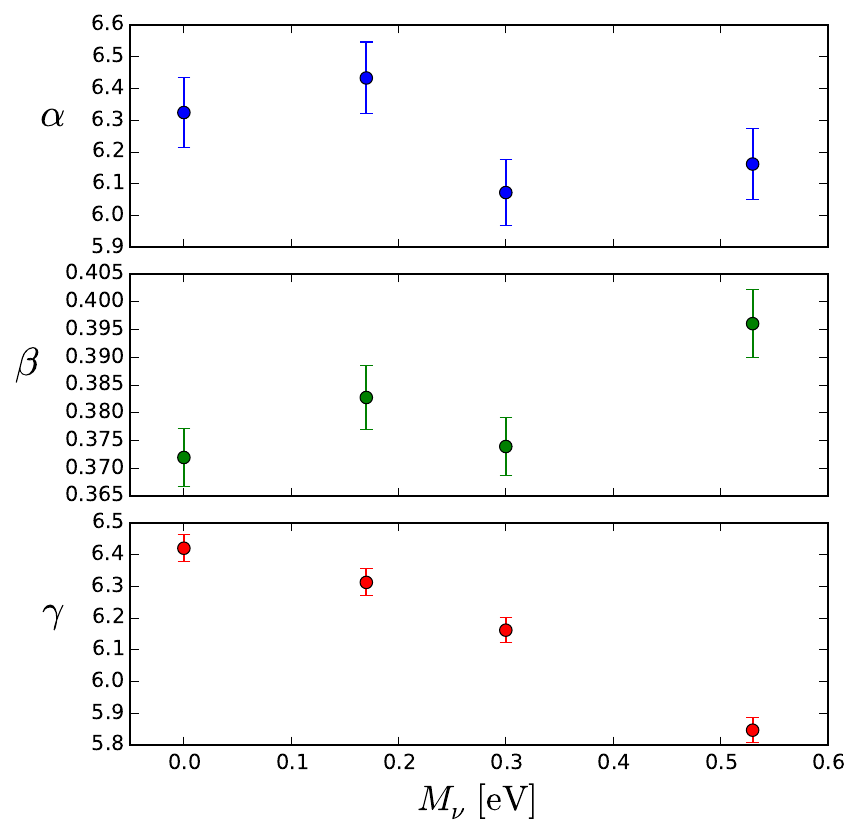}
\end{subfigure}
\caption{Same as figure \ref{fig_fit_summary_CDM_coreDens} for halo voids correlated with halos, bins in compensation~$\Delta_\halo$.}
\label{fig_fit_summary_Halos_deltaN}
\end{figure}

\begin{figure}[!ht]
\centering
\begin{subfigure}{.5\textwidth}
  \centering
  \includegraphics[width=.99\linewidth]{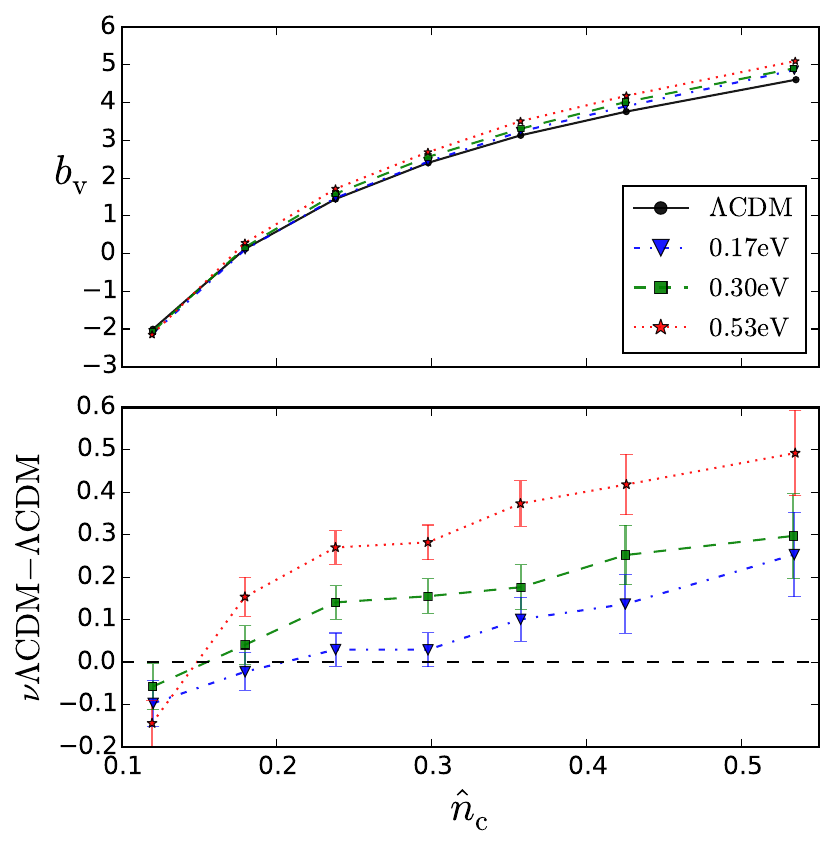}
\end{subfigure}%
\begin{subfigure}{.5\textwidth}
  \centering
  \includegraphics[width=.99\linewidth, trim = 10 8 0 10]{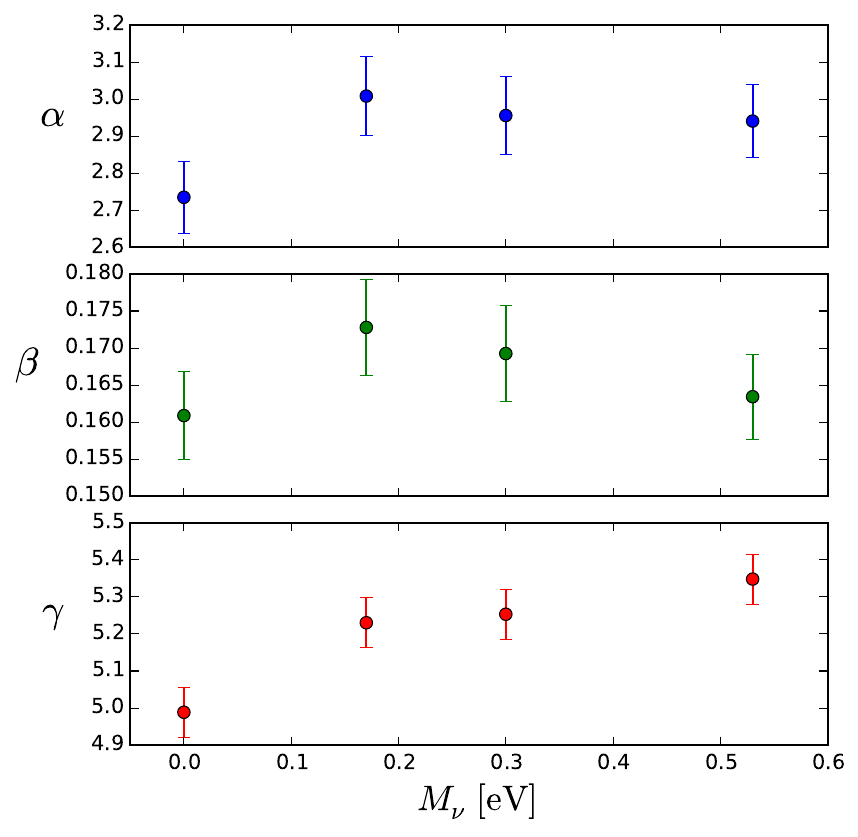}
\end{subfigure}
\caption{Same as figure \ref{fig_fit_summary_CDM_coreDens} for halo voids correlated with matter, bins in core density~$\coreDens$.}
\label{fig_fit_summary_HalosCDM_coreDens}
\end{figure}

\begin{figure}[!ht]
\centering
\begin{subfigure}{.5\textwidth}
  \centering
  \includegraphics[width=.99\linewidth]{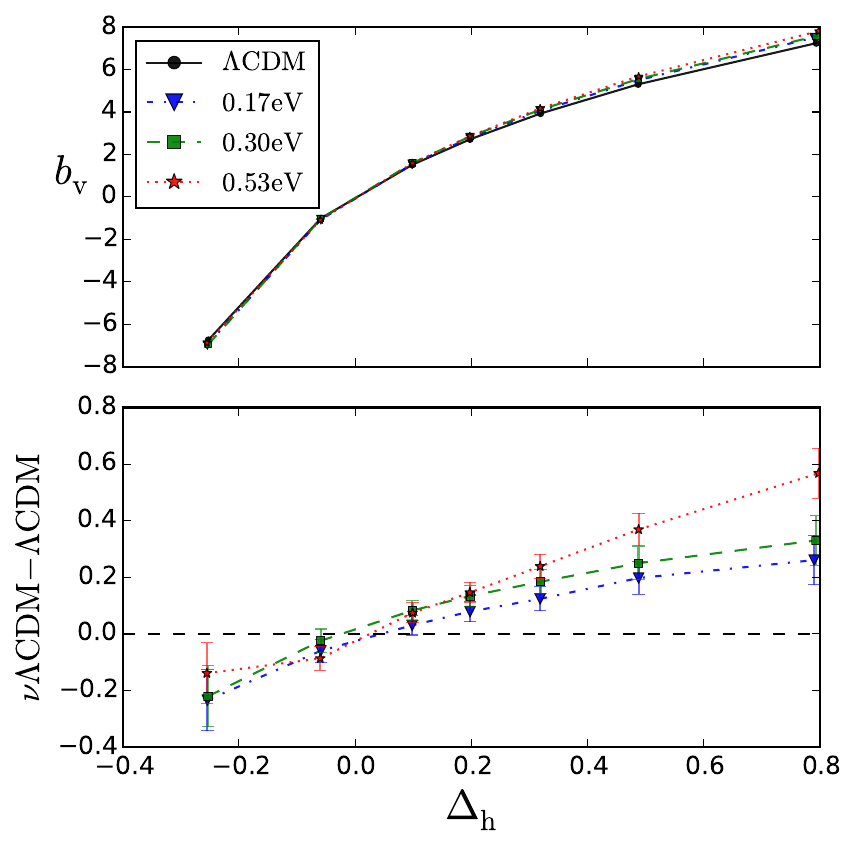}
\end{subfigure}%
\begin{subfigure}{.5\textwidth}
  \centering
  \includegraphics[width=.99\linewidth, trim = 0 12 0 9]{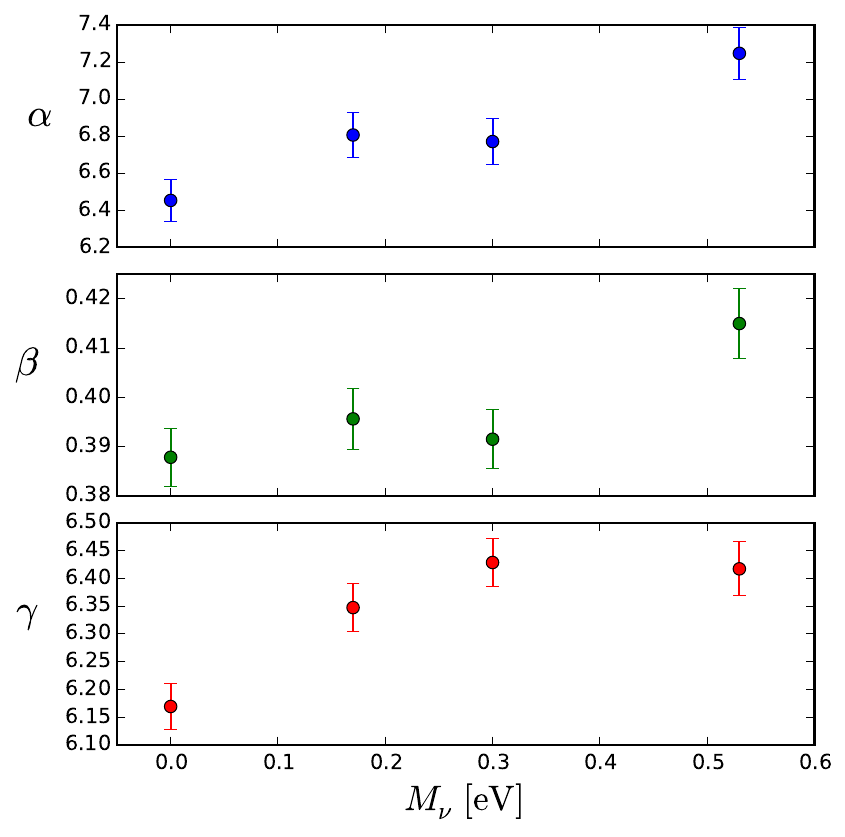}
\end{subfigure}
\caption{Same as figure \ref{fig_fit_summary_CDM_coreDens} for halo voids correlated with matter, bins in compensation~$\Delta_\halo$.}
\label{fig_fit_summary_HalosCDM_deltaN}
\end{figure}

Theoretical models in the literature that predict the void bias from first principles exist~\cite[e.g.,][]{Sheth2004,Furlanetto2006,Chan2014,Chan2018,Chan2019,Jamieson2019}, but they are still limited in their accuracy to describe the bias of halo voids. Moreover, massive neutrinos have so far not been included in that framework, although reference~\cite{Banerjee2016} presented a model for the bias of CDM voids including neutrinos. We therefore set out to find some function $f_b$ that can model the void bias in a phenomenological way. As variable for this unknown function we choose the void compensation $\Delta_\tracer$, because it correlates most strongly with $b_\void$, as evident from figures~\ref{fig_CDM_summary} and~\ref{fig_void_halos_deltaN_summary}. We find that the following parameterization provides a very good description of our numerical results:
\begin{equation}
    f_b(\Delta_\tracer) = \alpha \cdot \ln(\Delta_\tracer + \beta) + \gamma  \,,
    \label{eq:fit_function}
\end{equation}
where $\alpha, \beta$ and $\gamma$ are free parameters that are used in fitting this function to our measured values of the large-scale void bias. We perform the fit for CDM voids, as well as for halo voids. In the latter case we use either halos, or the total matter to correlate the voids with. In figure~\ref{fig_single_fits_CDM_coreDens} we present the fit of Eq.~(\ref{eq:fit_function}) to our simulation results, in this case for the CDM voids as a function of their core densities for the case $M_\nu = 0.53$eV. The plot shows the average void bias on large scales for our bins in $\coreDens$, with corresponding values of $\Delta_\matter$ shown on the $x$-axis. The best-fit function $f_b(\Delta_\matter$) is shown as a dotted line, with the shaded region indicating its error. We repeat this fitting process for each of our four neutrino simulations and for each void analysis. The results are presented in figures~\ref{fig_fit_summary_CDM_coreDens} and~\ref{fig_fit_summary_CDM_deltaN} for the CDM voids, in figures~\ref{fig_fit_summary_Halos_coreDens} and~\ref{fig_fit_summary_Halos_deltaN} for halo voids correlated with halos, and in figures~\ref{fig_fit_summary_HalosCDM_coreDens} and~\ref{fig_fit_summary_HalosCDM_deltaN} for halo voids correlated with the total matter, each case for the bins in core density and compensation. \AP{The lower left panels show the differences of the fits with respect to $\Lambda$CDM, and the right panels the corresponding best-fit parameters.}

In all the above cases the parameter $\gamma$ varies most significantly between different values of $M_\nu$. In particular, when CDM is used to define voids, or when matter is used to calculate the power spectra, $\gamma$ increases with $M_\nu$, but the opposite is true for halo voids when correlated with halos~\CK{\cite{Kreisch2018}}. A similar trend is evident in the amplitude parameter $\alpha$, albeit with more scatter between the values of $M_\nu$. Finally, $\beta$ is roughly consistent with a constant $x$-offset within the errors. We observe the function $f_b(\Delta_\tracer)$ to cross zero when $\Delta_\tracer\simeq0$, \AP{as expected for compensated voids~\cite{hamaus_voidPowSpec}}. The fits nicely mirror the behavior seen before, namely that the amplitude of the void bias increases when arranged in bins of void compensation.

\section{Conclusion\label{sec:conclusion}}

We have explored the influence of massive neutrinos on various void properties at redshift $z=0$, making use of the void finder \textsc{vide} and the \texttt{DEMNUni} simulation suite that incorporates sums of neutrino masses $\sumnu$ between $0.0$eV ($\Lambda$CDM) and $0.53$eV. We focused our analysis on how the abundance and clustering properties of voids are altered in massive neutrino cosmologies, with a particular focus on the large-scale void bias. \CK{Built upon the work by references~\cite{Massara2015,Pollina2016,Kreisch2018}, we explored how the chosen tracer distribution to identify voids (CDM or halos) affects our results}. We also distinguished voids of different core density and compensation, i.e. their average density of tracers compared to the background density. \EM{}\CK{}\AP{We chose to fix the number density of our tracer populations among all neutrino cosmologies, as opposed to performing a constant halo-mass cut. This eliminates differences in the abundance and clustering properties of voids which are purely caused by the varying sparsity of tracers with $M_\nu$. However, the halo-bias dependence on $M_\nu$ remains important for the properties of voids.}

We find that the total number of voids in the CDM distribution becomes larger with increasing neutrino mass~\CK{(cf.~\cite{Massara2015,Kreisch2018})}, and that the void abundance exhibits clear differences in the presence of massive neutrinos, both when plotted against core density and compensation. CDM voids tend to have higher core densities with increasing neutrino masses and are therefore shallower in $\nu \Lambda$CDM cosmologies~\cite{Massara2015}. The fraction of compensated CDM voids with $\Delta_\matter = 0$ increases in the massive neutrino simulations. For voids identified in the halo distribution the opposite happens~\CK{(cf.~\cite{Kreisch2018})}: total void numbers decrease with increasing neutrino mass, but the overall differences are much smaller than in the case of CDM voids. We also notice a small trend towards lower core densities, i.e. deeper and less compensated voids with $\Delta_\halo = 0$ in the $\nu \Lambda$CDM simulations. \CK{The inversion in the neutrino response between CDM voids and halo voids can be explained through the additional neutrino dependence of the halo bias~\cite{Kreisch2018}.}

We have further calculated various power spectra between voids and matter, as well as between voids and halos, and extracted the scale-dependent void bias $b_\void(k)$ via Eqs.~(\ref{eq:void_bias_CDM}) and~(\ref{eq:void_bias_halos}). As expected from earlier simulation results~\cite{hamaus_voidPowSpec,Chan2014}, the void bias attains a constant value on large scales (low $k$) and a dip, caused by the void-density profile, on smaller scales. The same overall trend is preserved in massive neutrinos cosmologies, but differences on both large and small scales appear. In the case of CDM voids the most striking effect is an increase in their large-scale bias amplitude with increasing $M_\nu$. Therefore, undercompensated voids with $\Delta_\matter<0$ and $b_\void<0$ tend to have a more negative $b_\void$ in $\nu$CDM, and conversely overcompensated voids with $\Delta_\matter>0$ and $b_\void>0$ a more positive $b_\void$. This implies that exactly compensated voids with $\Delta_\matter=0$ and $b_\void=0$ feature no neutrino sensitivity in the void bias on large scales, which we confirm with our simulation results. However, further scale-dependent signatures in $b_\void(k)$ from neutrinos on smaller scales are present as well. The latter can be helpful to break degeneracies with other cosmological parameters, such as the normalization of the power spectrum, $\sigma_8$.

In our simulations $\sigma_8$ varies with $M_\nu$, since the amplitude of primordial curvature perturbations $A_s$ is held constant. The degree of degeneracy between $M_\nu$ and $\sigma_8$ in the void bias $b_\void$ remains to be investigated in simulations with fixed $\sigma_8$ values. \EM{\AP{However, reference~\cite{Massara2015} already found the $M_\nu-\sigma_8$ degeneracy to be broken in the abundance, density and velocity profiles of CDM voids in simulations.} Therefore, any such degeneracy for the void bias, which could potentially affect neutrino constraints, is not expected to be severe.}

When switching to halo voids to compute the relative bias between voids and halos as an observable quantity~(cf.~\cite{Pollina2018}), we again notice an inversion of the previous behavior~\CK{\cite{Kreisch2018}} with generally lower amplitude. As before, this has to do with the halo-bias sensitivity to the neutrino mass, which counteracts the neutrino imprints in voids~\CK{\cite{Kreisch2018}}. However, when using the total matter distribution to correlate with the halo voids, the original behaviour observed for CDM voids can be restored, and the amplitude of the neutrino imprints are enhanced. The latter case is relevant for observations that combine weak lensing and redshift survey data. Note, however, that in this case the line-of-sight resolution of the reconstructed matter density field is very low (which is why we cannot access voids in the 3D matter field in the first place), so one needs to consider projected power spectra in a more realistic analysis.

Finally, we have investigated a phenomenological model to describe the void bias under the influence of massive neutrinos. Eq.~(\ref{eq:fit_function}) manages to reproduce our data remarkably well, and two out of its three free parameters exhibit significant sensitivity to the sum of neutrino masses. Despite these results, a theoretical model predicting the void bias from first principles is needed and will be investigated in future work. The void bias has already been measured from void correlation functions in SDSS~\cite{Clampitt2016}, ranging from positive to negative values as the void size increases, in agreement with simulations~\cite{hamaus_voidPowSpec,Chan2014}. It will be exciting to explore similar data in the context of massive neutrinos with the upcoming surveys, such as DESI, Euclid, LSST, or WFIRST~\cite{Pisani2019}. These will have access to much larger volumes of large-scale structure and thus provide significantly higher constraining power than what is presented in this work.

\begin{acknowledgments}
We thank Elena Massara, David Spergel, Matteo Viel, Jia Liu, Ben Wandelt, and Uro\v s Seljak for fruitful discussions and comments. NS, NH, GP and JW acknowledge support from the DFG cluster of excellence ``Origins'' and the Trans-Regional Collaborative Research Center TRR 33 ``The Dark Universe'' of the DFG. AP is supported by NASA grant 15-WFIRST15-0008 to the WFIRST Science Investigation Team ``Cosmology with the High Latitude Survey''. CC acknowledges support by the Advanced Research Grant of the European Research Council (n. 291521) and from ASI through grant I/023/12/0. The DEMNUni-I simulations were carried out in the framework of the ``The Dark Energy and Massive-Neutrino Universe" using the Tier-0 IBM BG/Q Fermi machine of the Centro Interuniversitario del Nord-Est per il Calcolo Elettronico (CINECA). We acknowledge a generous CPU and storage allocation by the Italian Super-Computing Resource Allocation (ISCRA). CDK is supported by the National Science Foundation Graduate Research Fellowship under Grant DGE 1656466.
\end{acknowledgments}

\bibliography{ms,msHamaus}

\bibliographystyle{JHEP}

\end{document}